\documentclass[10pt,preprint,prd,eqsecnum,floatfix]{revtex4}
\usepackage{amsmath,graphicx}
\begin{document}
%-----------------------------------------------------------------------------

\title{Hydrodynamic Stability of Cosmological 
       Quark-Hadron Phase Transitions}

\author{P. Chris Fragile and Peter Anninos}
\affiliation{University of California,
Lawrence Livermore National Laboratory, Livermore, CA 94550}

%\date{{\small    \today}}
%\date{{\small   \LaTeX-ed \today}}
%-----------------------------------------------------------------------------

\begin{abstract}
Recent results from linear perturbation theory suggest that
first-order cosmological quark-hadron phase transitions occurring
as deflagrations may be ``borderline'' unstable, and
those occurring as detonations may give rise to growing modes behind the
interface boundary. However, since nonlinear effects can play important
roles in the development of perturbations, 
unstable behavior cannot be asserted entirely by linear analysis, and 
the uncertainty of these recent studies is compounded further by 
nonlinearities in the hydrodynamics and self-interaction
fields. In this paper we investigate the growth
of perturbations and the stability of 
quark-hadron phase transitions
in the early Universe by solving numerically
the fully nonlinear relativistic hydrodynamics equations
coupled to a scalar field with a quartic self-interaction
potential regulating the transitions. We consider single, perturbed,
phase transitions propagating either by detonation or deflagration,
as well as multiple phase and shock front interactions in 
1+2 dimensional spacetimes.

\vspace*{5mm} 
\end{abstract}

\maketitle
%-----------------------------------------------------------------------------

\section{Introduction}
\setcounter{equation}{0}
\label{sec:intro}

First-order phase transitions ocurring at either the electroweak
or QCD symmetry breaking epochs in the early Universe, as predicted
by the standard model of cosmology,
can have important consequences for the history of
our Universe. In particular, the formation of 
co-existing bubbles and droplets of different phases during the QCD transition
affects the production and distribution of
hadrons, and may lead to significant baryon number density 
fluctuations. 
Assuming these fluctuations survive to the epoch of
Big Bang nucleosynthesis, they can strongly influence the predicted
light element abundances \cite{FMA88}.
This, in turn, may alter our current understanding
and interpretation of homogeneous Big Bang nucleosynthesis,
as well as the dynamical and chemical
evolution of matter structures, including the origins of
primordial magnetic fields and structure
perturbations that seed the subsequent formation of
stellar, galactic, and cluster scale systems.

Baryon density fluctuations evolve hydrodynamically by the competing
effects of local phase mixing and phase separation that
may arise during the transition period. A potential mixing mechanism
that we consider here 
is the hydrodynamic instability of the interface surfaces
separating regions of different phase. Unstable modes may
distort the phase boundary and
induce mixing and diffusive homogenization
through hydrodynamic turbulence.
Whether unstable modes can exist in
cosmological phase fronts has been discussed 
recently by several authors with conflicting conclusions 
based on first-order perturbation theory.
For example, Kamionkowski and Freese \cite{KF92}
suggest that subsonic deflagration fronts in electroweak transitions
may be accelerated until they become supersonic and proceed
as detonations through an effective increase
in the front velocity due to surface distortion effects as
the transition becomes turbulent.
Link \cite{Link92} indicated that the phase boundary in slow
deflagrations from quark-hadron transitions may be unstable to
long wavelength perturbations, but may be stabilized by
surface tension at shorter wavelengths.
In contrast, Huet et al. \cite{HKLLM93} find that for electroweak
transitions, the shape of the phase boundary is stable under
perturbations, and quark-hadron transitions are at the ``borderline''
between stable and unstable. They also argue that unstable
modes are not possible in detonation fronts. However,
Abney \cite{Abney94} suggests that this is the case
for the quark phase ahead of the supersonic interface boundary, 
but that the cold
phase hadron regions behind the bubbles might be unstable,
at least for the Chapman-Jouget type detonations
investigated.

These results are certainly not conclusive,
and in some cases are in apparent contradiction due
to the level of approximation and coupling 
assumed between thermal and dynamical variables.
Also, the full consequences or even existence
of instabilities cannot be determined entirely by
linear analysis. Nonlinear effects, including 
higher-order coupling between hydrodynamic, microphysical,
scalar field, and interaction potentials, as well as
surface tension, entropy flow,
and baryon transport, may all play important
roles in the stabilization (or de-stabilization) of
phase boundaries, and remain to be investigated in detail.
Hence, we undertake this study to more fully investigate the hydrodynamic
stability of cosmological phase transitions ocurring during the QCD epoch. 
We take a numerical approach and thus generalize previous
results by solving the multi-dimensional
relativistic hydrodynamics equations, presented
in \S\ref{sec:equations}, coupled together with a model
scalar wave equation and an interaction potential
regulating the phase transitions. Using results from
perturbation theory to define initial data (as discussed
in \S\ref{sec:perturbation}), we solve numerically for the evolutions
of single distorted phase fronts as well as interactions and
collisions of multiple front systems propagating as either
deflagrations or detonations. Our numerical results are
presented in \S\ref{sec:results}, and summarized in \S\ref{sec:summary}.

\section{Basic Equations}
\setcounter{equation}{0}
\label{sec:equations}

The equations formulated with internal fluid energy are derived
from the 4-velocity normalization $u^{\mu}u_{\mu} = -1$,
the parallel component of the stress--energy conservation equation
$u_\nu\nabla_{\mu} T^{\mu\nu} = 0$ for internal energy,
the transverse component of the stress--energy equation
$(g_{\alpha\nu} + u_\alpha u_\nu) \nabla_{\mu} T^{\mu\nu} = 0$ for momentum,
and an equation of state 
for the fluid pressure and temperature, using traditional
high energy units in which $c=\hbar=k_B=1$.
We consider the following special relativistic stress-energy tensor
\begin{equation}
T^{\mu\nu} = \rho h u^\mu u^\nu + p_r g^{\mu\nu}
           + \partial^\mu\phi \partial^\nu\phi
           - g^{\mu\nu}\left(\frac12 \partial_\alpha\phi \partial^\alpha\phi
                             + V(\phi,~T)\right) ~,
\label{eqn:tmn}
\end{equation}
which includes both fluid and scalar field ($\phi$) contributions.
$\rho h$ is the relativistic enthalpy,
$p_r$ is the radiation pressure,
$u^\mu$ is the contravariant 4-velocity, and
$g_{\mu\nu}=\eta_{\mu\nu}$ is the flat space 4-metric. 

\begin{figure}
\includegraphics[width=10.5cm]{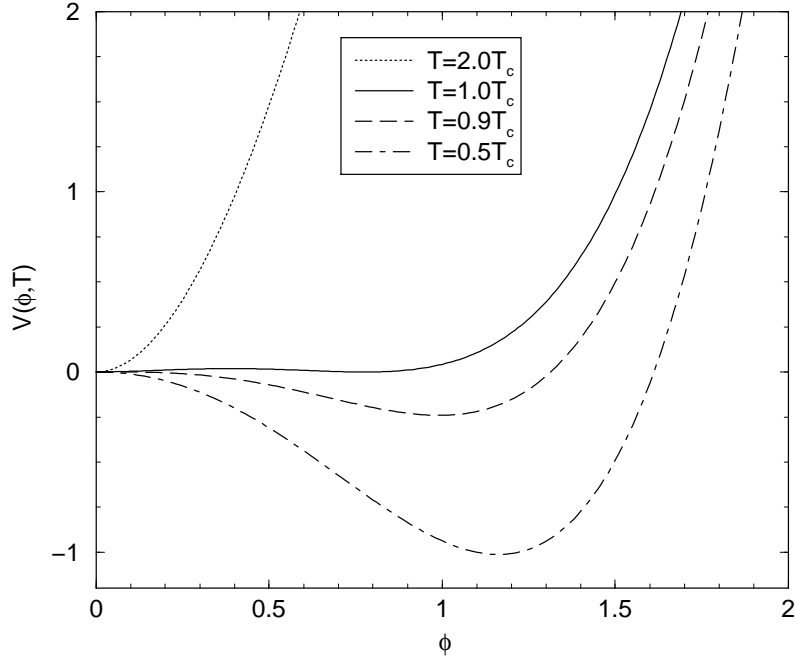}
\caption{Interaction potential $V(\phi,~T)$ 
(equation (\protect{\ref{eqn:pot}}) in the text) as a function of
scalar field at four different temperatures 
$T/T_c = (2,~1,~0.9,~0.5)$. The potential parameters
used here are:
surface tension $\sigma=0.1~T_c$, 
correlation length $\ell_c=1.0/T_c$,
latent heat $L=2.0~T_c^4$, 
and critical temperature $T_c=1$.
\label{fig:pot}}
\end{figure}

For the
scalar field self-interaction potential $V(\phi,~T)$, we adopt the form
\cite{Linde83}
\begin{equation}
V(\phi,~T) = \frac14\lambda\phi^4 - \frac13\alpha T \phi^3 
          + \frac12\beta(T^2 - T_0^2)\phi^2 ~,
\label{eqn:pot}
\end{equation}
where $T$ is the fluid temperature, and
$\lambda$, $\alpha$, $\beta$, and $T_0$ are constant parameters
specified indirectly through their relationships to the
critical transition temperature $T_c$, surface
tension $\sigma$, correlation length $\ell_c$, and latent heat $L$
\cite{Kajantie92,EIKR92}
\begin{equation}
\left\{ \begin{array}{ll}
        T_0~, & \alpha \\
        \beta~, & \lambda
        \end{array}
\right\}
=
\left\{ \begin{array}{ll}
        T_c [1+6\sigma/(L\ell_c)]^{-1/2}~, & (2\sigma\ell_c^5 T_c^2/3)^{-1/2} \\
        (L+6\sigma/\ell_c)(6\sigma\ell_c T_c^2)^{-1}~, & (3\sigma\ell_c^3)^{-1}
        \end{array}
\right\} .
\end{equation}
The potential (\ref{eqn:pot}) is plotted at four
different temperatures $T/T_c = (2,~1,~0.9,~0.5)$ as a function of
scalar field $\phi$ in Figure \ref{fig:pot}.  At high temperatures ($T>T_c$) 
the potential exhibits a single minimum ($\phi=V(\phi,~T)=0$), 
where the cosmological fluid is essentially a
quark plasma of unbroken symmetry. 
A second minimum ($\phi=\phi_{min}(T)$), corresponding to a low temperature
hadron phase, occurs at temperatures between $T_0$ and $T_c$, where
\begin{equation}
\phi_{min}(T) = \frac{\alpha T}{2\lambda} 
          \left(1+\sqrt{1-\frac{4\lambda\beta(T^2-T_0^2)}{\alpha^2 T^2}}\right) ~.
\end{equation}
Below $T_0$ the potential has a local maximum at $\phi=0$ and a single 
minimum at $\phi=\phi_{min}(T)$.

The resulting differential equations for momentum and energy, neglecting baryon
conservation, can be written in flux conservative form as
\begin{eqnarray}
\partial_t S_j + {\partial_i}(S_j v^i) &=&
    - \partial_j [p_r-V(\phi,~T)] 
    - \left[\kappa W \partial_t \phi + \kappa W v^i \partial_i \phi 
    + \partial_\phi V(\phi,~T)\right] \partial_j\phi ~, \label{eqn:mom} \\
\partial_t E + {\partial_i}(E v^i) &=&
    - [p_r-V(\phi,~T)] [\partial_t W + {\partial_i}(W v^i) ] \nonumber \\
&&  + \kappa (W \partial_t \phi + W v^i \partial_i \phi)^2
    + \partial_\phi V(\phi,~T) (W \partial_t \phi + W v^i \partial_i \phi)~, 
    \label{eqn:en1}
\end{eqnarray}
which, together with the scalar field wave equation,
\begin{equation}
\partial^2_t \phi - \partial^i \partial_i \phi =
    -\partial_\phi V(\phi,~T) 
    - \kappa W (\partial_t\phi + v^i\partial_i\phi)~,
\label{eqn:scalar}
\end{equation}
and a prescription for the equation of state described below,
represent the complete system of equations we solve in this paper.
In deriving the above equations, we
have implicitly assumed a flat space metric,
and defined variables so that
$W=u^0=1/\sqrt{1-v^i v_i}$ is the relativistic boost factor, 
$v^i=u^i/u^0$ is the transport velocity,
$S_i = W\rho h u_i$ is the covariant momentum density, 
$E$ is the generalized internal energy density 
\begin{equation}
\frac{E}{W} = 3 a_r T^4 + V(\phi,~T) - T \partial_T V(\phi,~T)~,
\end{equation}
$a_r=g_*(\pi^2/90)$ is a radiation constant defining the particle
content and degrees of freedom $g_*$,
and $\kappa$ is a constant coefficient regulating entropy
production and kinematic and scalar field energy dissipation.
The terms associated with $\kappa$ act
as an effective friction force at the phase
transition surface boundary \cite{IKKL94,KL96}.

The hydrodynamic equations are solved using the Cosmos code
\cite{AF02,AFM02}
with time-explicit operator split methods,
second order spatial finite differencing, and artificial
viscosity to capture shocks \cite{Wilson79}.
The scalar field wave equation is solved using a
second order (in space and time) predictor-corrector 
integration, sub-cycling within a single 
hydrodynamic cycle since equation (\ref{eqn:scalar}) generally
has a tighter restriction on the time step for stability.

A second formulation of the dynamical equations considered
here is based on a simpler conservative hyperbolic 
description of the hydrodynamics equations. In this case,
the equations are derived directly by
expanding the conservation of stress-energy
$T^{\mu\nu}_{\ \ ,\nu} = 0$ into time and space explicit parts.
In flat space, this yields the same equations
(\ref{eqn:mom}) and (\ref{eqn:scalar}) for momentum conservation
and scalar field evolution, but with
\begin{equation}
\partial_t {\cal E} + {\partial_i} ({\cal E}v^i) =
    - {\partial_i} [v^i(p_r-V(\phi,~T))] 
    + \left[\kappa W \partial_t \phi + \kappa W v^i \partial_i \phi
    + \partial_\phi V(\phi,~T)\right] \partial_t \phi 
\label{eqn:en2}
\end{equation}
in place of (\ref{eqn:en1}) for the energy variable,
defined as
\begin{equation}
{\cal E}   = W \rho h u^0 - [p_r-V(\phi,~T)] .
\end{equation}
The Cosmos code uses a non-oscillatory central 
difference numerical scheme \cite{JLLOT98}
with second order spatial differencing and
predictor-corrector time integration
with dimensional splitting
to solve the hydrodynamics equations in this formalism.
Artificial viscosity is not needed in this approach. Instead,
shocks are captured with MUSCL-type piece-wise linear interpolants
and high order limiters for flux reconstruction, but without the
complexity of Riemann solvers.

A final ingredient needed is an equation of state
relating $p_r$ to energy $E$ and temperature $T$.
Here we assume $p_r = a_r T^4$, treating the
plasma as a gas of relativistic particles with enthalpy density
\begin{equation}
\rho h = \frac{\gamma p_r}{\gamma-1} - T\partial_T V(\phi,~T)
       = 4 a_r T^4 - T \partial_T V(\phi,~T)
       = \frac{E}{W} + p_r-V(\phi,~T) ~,
\end{equation}
and adiabatic index $\gamma=4/3$.
The equation of state is then written in terms of computed quantities
\begin{equation}
p_{eff} = p_r - V(\phi,~T) = (\gamma-1)\left(\frac{E}{W}
                      + T\partial_T V(\phi,~T) - V(\phi,~T) \right) - V(\phi,~T) ~,
\label{eqn:eos1}
\end{equation}
or
\begin{equation}
p_{eff} = p_r - V(\phi,~T) = 
   (\gamma-1)\left(\frac{{\cal E} - V(\phi,~T) + TW^2\partial_T V(\phi,~T)}
                        {\gamma W^2 - (\gamma-1)}\right) - V(\phi,~T) ~,
\label{eqn:eos2}
\end{equation}
depending on which energy variable is evolved.
The quantity $p_{eff}$ is introduced as an effective pressure
for numerical convenience in solving equations (\ref{eqn:mom}),
(\ref{eqn:en1}) and/or (\ref{eqn:en2}).
The fluid temperature is computed from either
\begin{equation}
\frac{E}{W} - 3 a_r T^4 - V(\phi,~T) + T\partial_T V(\phi,~T) = 0 ~, 
\label{eqn:temp1}
\end{equation}
or 
\begin{equation}
{\cal E} - V(\phi,~T) + TW^2\partial_T V(\phi,~T)
    - 3 a_r T^4 [\gamma W^2 - (\gamma-1)] = 0 ~.
\label{eqn:temp2}
\end{equation}
Equations (\ref{eqn:temp1}) and (\ref{eqn:temp2}) are solved
using a Newton-Raphson method to iterate and converge on the
temperature, assuming an initial guess of
$T_{0} = (E/3Wa_r)^{1/4}$.

\section{Perturbation Analysis}
\setcounter{equation}{0}
\label{sec:perturbation}

In this section we carry out a first-order perturbation expansion
of the hydrodynamics equations, and review briefly the expected
dynamical behavior and stability of phase transitions in the
context of linear theory. The general aproach follows that
presented in references \cite{Link92,KF92,HKLLM93,Abney94}. 
In addition to elucidating
any unstable behavior, the results of this section are also
used to set up inhomogeneous perturbations as initial data
that is evolved numerically in \S\ref{sec:results}.

It is convenient to start with the hydrodynamics equations in
conservative hyperbolic form (\ref{eqn:mom}) and
(\ref{eqn:en2}), which we rewrite here as
\begin{eqnarray}
\partial_t {\cal E} + \partial_i ({\cal E} v^i + v^i p) 
       &=& \Sigma^0 ~, \label{eqn:en_pert} \\
\partial_t S^j + \partial_i (S^j v^i + \delta_{ij}  p) 
       &=& \Sigma^j ~, \label{eqn:mom_pert}
\end{eqnarray}
assuming a constant scalar field potential on either side
of the phase transition front, and
$\Sigma^0$ and $\Sigma^j$ represent any source or dissipative terms.
Equations (\ref{eqn:en_pert}) and (\ref{eqn:mom_pert}) are expanded
out to first perturbative order assuming 
two dimensional perturbations off a homogeneous background
flow of the form
\begin{eqnarray}
p &=& p_0 + \delta p(t,x,y) ~, \\
{\bf v} &=& v_0~\widehat i 
            + [\delta v_x(t,x,y)~\widehat i + 
               \delta v_y(t,x,y)~\widehat j] ~,
\end{eqnarray}
for the fluid pressure and velocity.
The corresponding boost factor and conserved variables 
are, to first order
\begin{eqnarray}
W &=& \frac{1}{\sqrt{1-v^2}}
   =  \frac{1}{\sqrt{1-v_0^2}} + \frac{v_0}{(1-v_0^2)^{3/2}}~\delta v_x ~,
   \label{eqn:boost_pert} \\
{\cal E} &=& \frac{p(\gamma W^2 - \gamma + 1)}{\gamma-1}
   = -p_0 + \frac{\gamma p_0}{(\gamma-1)(1-v_0^2)} \nonumber \\
   && \hskip85pt
     + \frac{2\gamma p_0 v_0}{(\gamma-1)(1-v_0^2)^2}~\delta v_x
     + \frac{\gamma}{(\gamma-1)(1-v_0^2)}~\delta p - \delta p ~, \\
S^x &=& \frac{\rho h v^x}{1-v^2}
     = \frac{\gamma}{(\gamma-1)(1-v_0^2)}
       \left(p_0 v_0 + p_0 ~\delta v_x + v_0~\delta p + 
            \frac{2 p_0 v_0^2}{1-v_0^2}~\delta v_x\right) ~, \\
S^y &=& \frac{\rho h v^y}{1-v^2}
     = \frac{\gamma p_0}{(\gamma-1)(1-v_0^2)}~\delta v_y ~,
       \label{eqn:sy_pert}
\end{eqnarray}
where $\rho h = (e+p) = \gamma p/(\gamma-1)$, and $v_0$
is the unperturbed fluid velocity.

Substituting (\ref{eqn:boost_pert}) - (\ref{eqn:sy_pert})
into (\ref{eqn:en_pert}) and (\ref{eqn:mom_pert}) yields
for momentum conservation along the $y$-axis
\begin{equation}
\partial_y (\delta p) + \frac{\gamma p_0}{(\gamma-1)(1-v_0^2)} 
      \left[\partial_t (\delta v_y) + v_0 \partial_x (\delta v_y)
      \right] = \Sigma^y ~,
\label{eqn:sigmay}
\end{equation}
for momentum conservation along the $x$-axis
\begin{eqnarray}
(\gamma-1)(1-v_0^2) \Sigma^x 
&=&\gamma v_0~\partial_t (\delta p) 
+\gamma p_0(1+v_0^2)(1-v_0^2)^{-1}~\partial_t (\delta v_x)
\nonumber \\
&+&(\gamma-1 + v_0^2)~\partial_x (\delta p)
+2\gamma p_0 v_0 (1-v_0^2)^{-1}~\partial_x (\delta v_x) 
+\gamma p_0 v_0~\partial_y (\delta v_y)
\label{eqn:sigmax} ~,
\end{eqnarray}
and for energy conservation
\begin{eqnarray}
(\gamma-1)(1-v_0^2) \Sigma^0
&=&\left(1-v_0^2+\gamma v_0^2\right)~\partial_t (\delta p)
+2\gamma p_0 v_0 (1-v_0^2)^{-1}~\partial_t (\delta v_x)
\nonumber \\
&+&\gamma v_0~\partial_x (\delta p) 
+\gamma p_0 (1+v_0^2)(1-v_0^2)^{-1}~\partial_x (\delta v_x)
+\gamma p_0~\partial_y (\delta v_y) ~,
\label{eqn:sigma0}
\end{eqnarray}
all to first perturbative order.

Equations (\ref{eqn:sigmax}) and (\ref{eqn:sigma0})
can be simplified further by first eliminating
$\partial_y (\delta v_y)$ from the $x$-momentum equation
\begin{eqnarray}
0 &=& (\gamma-1)(1-v_0^2)\left(\Sigma^x - v_0 \Sigma^0\right) \nonumber \\
&=&
v_0 \partial_t (\delta p) + \rho h_0 W_0^2 \partial_t (\delta v_x)
+ \partial_x (\delta p) + W_0^2 \rho h_0 v_0 \partial_x (\delta v_x) ~,
\label{eqn:sigmax2}
\end{eqnarray}
then eliminating $\partial_t (\delta v_x)$ from the energy equation
\begin{eqnarray}
0 &=& c_s^2(\gamma-1)(1-v_0^2)
      \left[\Sigma^0 - \left(\frac{2v_0}{1+v_0^2}\right) \Sigma^x\right]
  \nonumber \\
&=& 
(1 - v_0^2 c_s^2)\partial_t (\delta p) + v_0(1 -c_s^2)\partial_x (\delta p)
+ \rho h_0 \left[\partial_x (\delta v_x) + \partial_y (\delta v_y)\right] ~.
\label{eqn:sigma02}
\end{eqnarray}
In writing (\ref{eqn:sigmax2}) and (\ref{eqn:sigma02}) we
introduced the notation $c_s^{2}=\gamma-1$, $W_0^2=1/(1-v_0^2)$
$\rho h_0 = \gamma p_0/(\gamma-1)$, and
explicitly ignored the source terms 
$\Sigma^\alpha$ by setting them to zero.

The final equations (\ref{eqn:sigmay}), (\ref{eqn:sigmax2}), and
(\ref{eqn:sigma02}) can be written conveniently in compact form as
\begin{equation}
{\bf A_t} \partial_t {\bf W} + {\bf A_x} \partial_x {\bf W} 
+ {\bf A_y} \partial_y {\bf W} = 0 ~,
\label{eqn:compact}
\end{equation}
where ${\bf W} = (\delta p,~\delta v_x,~\delta v_y)$, and
\begin{eqnarray}
{\bf A_t} &=& \left(\begin{array}{lll}
1-v_0^2 c_s^2 & 0 & 0 \\
v_0 & \rho h_0 W_0^2 & 0 \\
0 & 0 & \rho h_0 W_0^2 
                  \end{array}
            \right) ~,
\\
{\bf A_x} &=& \left(\begin{array}{lll}
v_0(1-c_s^2) & \rho h_0 c_s^2 & 0 \\
1 & W_0^2 \rho h_0 v_0 & 0 \\
0 & 0 & \rho h_0 W_0^2 v_0
                  \end{array}
            \right) ~,
\\
{\bf A_y} &=& \left(\begin{array}{lll}
0 & 0 & \rho h_0 c_s^2 \\
0 & 0 & 0 \\
1 & 0 & 0
                  \end{array}
            \right) ~.
\end{eqnarray}
Next, following the general procedure outlined
in reference \cite{Abney94}, we assume a solution of the form
${\bf W}(t,x,y) = {\bf w}(x) e^{-i(\omega t + k y)}$, with
${\bf w}(x) = \sum a_j e^{-i\lambda^*_j x} {\bf R_j}$ 
and the following characteristic equation or dispersion relation 
for (\ref{eqn:compact}):
\begin{equation}
(\omega + \lambda^* v_0)[k^2(1-v_0^2) - c_s^{-2}(\omega+\lambda^* v_0)^2
      + (\lambda^* + \omega v_0)^2] = 0 ~.
\label{eqn:char}
\end{equation}
Equation (\ref{eqn:char}) has either three distinct roots
which agree with those found by \cite{HKLLM93,Abney94}
\begin{eqnarray}
\lambda^*_1 &=& -\frac{\omega}{v_0} ~, \\
\lambda^*_{\pm} &=& \frac{1}{v_0^2-c_s^2}\left[(c_s^2-1)v_0 \omega \pm
c_s(1-v_0^2)\sqrt{\omega^2+\frac{v_0^2-c_s^2}{1-v_0^2}~k^2}\right] ~,
\end{eqnarray}
or two roots
\begin{eqnarray}
\lambda^*_1 &=& -\frac{\omega}{c_s} ~, \\
\lambda^*_{2} &=& \frac{v_0 k^2}{2\omega}
            - \frac{\omega(1+v_0^2)}{2v_0} ~,
\end{eqnarray}
if $v_0=c_s$. The corresponding eigenvectors of 
(\ref{eqn:compact}) are
\begin{eqnarray}
{\bf R_1} &=& \left(\begin{array}{l}
0 \\
k v_0/\omega \\
1
              \end{array}
            \right) ~,
\\
{\bf R_-} &=& \left(\begin{array}{l}
\left(\rho h_0\omega c_s^2 + \rho h_0 v_0 c_s \Omega \right)
     /(k(v_0^2-c_s^2)) \\
\left(v_0\omega + c_s \Omega\right)(1-v_0^2)
     /(k(c_s^2-v_0^2)) \\
1
              \end{array}
        \right) ~,
\label{eqn:rm}
\\
{\bf R_+} &=& \left(\begin{array}{l}
\left(\rho h_0 \omega c_s^2 - \rho h_0 v_0 c_s \Omega\right)
     /(k(v_0^2-c_s^2)) \\
\left(v_0\omega - c_s \Omega\right)(1-v_0^2)
     /(k(c_s^2-v_0^2)) \\
1
              \end{array}
        \right) ~,
\label{eqn:rp}
\end{eqnarray}
where we define $\Omega = [\omega^2+k^2(v_0^2-c_s^2)/(1-v_0^2)]^{1/2}$
in equations (\ref{eqn:rm}) and (\ref{eqn:rp}).

To determine whether any unstable modes of ${\bf W}$ exist
and are consistent with imposed boundedness conditions
at $x \rightarrow \pm \infty$, it is convenient to
construct a coordinate system centered and moving with
the frame of the interface between quark and hadron phases
at $x=0$. Assuming the high temperature quark phase is to the left 
of the interface ($x<0$), and the hadron phase to
the right ($x>0$), we can easily determine
if any unstable modes (defined by Im$(\omega)>0$)
obey the conditions ${\bf W}\rightarrow 0$ as
$x\rightarrow \pm \infty$. As concluded in references
\cite{HKLLM93,Abney94}, if Im$(\omega)>0$ in the quark region
of a detonation front,
then Im$(\lambda_j^*)<0$ and therefore
$a_j=0$ for all $j$
in order for the solutions
to be bounded at $x\rightarrow -\infty$. Hence
unstable modes do not exist in regions ahead of nucleating
bubbles separated from the quark phase by a detonation front.
However, in the hadron region to the right ($x>0$) of
a detonation front,
the boundedness conditions at $x\rightarrow +\infty$
do not rule out unstable modes to linear order. Also, in the case
of deflagration fronts, unstable modes are 
realizable on both the quark and hadron sides, at least for
cases in which surface tension, dissipation,
and strong thermal coupling effects
are negligible \cite{Link92,HKLLM93,KF92}.
It is not entirely clear what role these effects
play in stabilizing the transition, and certainly
higher order nonlinear effects are not accounted for in this analysis.
Table \ref{tab:regions} summarizes which of the regions and hydrodynamic
states may potentially give rise to unstable modes with
$\mbox{Im}(\omega)>0$, at least to linear order.

\begin{table}
\caption{
Summary of parameters yielding bound solutions that are
susceptible to unstable modes.  The entries represent a general
set of conditions that allow unstable modes ($\mbox{Im}(\omega)>0$)
to first perturbative order for each of the phases and front types.
These conditions are used to construct initial
data for the multi-dimensional numerical stability studies presented
in section \S\protect{\ref{sec:results}}.
\label{tab:regions}}
\begin{ruledtabular}
\begin{tabular}{ll}
Quark region ($x<0$)	&	Hadron region ($x>0$) \\
detonations:         	&	detonations:  \\
\qquad $v_q > c_s$		& 	\qquad $v_h < v_q$  \\
\qquad $a_1 = 0$		&	\qquad $\lambda^*_1 = -\omega/v_0$  \\
\qquad $a_- = 0$		&	\qquad $a_- = 0$ \\
\qquad $a_+ = 0$		&	\qquad Im$(\lambda^*_+) < 0$  \\
deflagrations:          &	deflagrations:  \\
\qquad $v_q < c_s$		& 	\qquad $v_h > v_q$  \\
\qquad $a_1 = 0$		&	\qquad $\lambda^*_1 = -\omega/v_0$  \\
\qquad Im$(\lambda^*_-) > 0$	&	\qquad $a_- = 0$ \\
\qquad $a_+ = 0$		&	\qquad Im$(\lambda^*_+) < 0$  \\
\end{tabular}
\end{ruledtabular}
\end{table}

%\vskip1pt
%\begin{equation}
%\begin{array}{l|l}
%\hline 
%\hline 
%\mbox{\bf quark region}~(x<0)          	& \mbox{\bf hadron region}~(x>0) \\
%\mbox{detonations:} \dotfill           	& \dotfill \\
%  \qquad v_q > c_s           	       	& v_h < v_q  \\
%  \qquad \mbox{Im}(\omega) > 0         	& \mbox{Im}(\omega) > 0    \\
%  \qquad \lambda^*_1 = -\omega/v_0 	& \lambda^*_1 = -\omega/v_0    \\
%  \qquad a_- = 0    	             	& \mbox{Im}(\lambda^*_-) < 0    \\
%  \qquad \mbox{Im}(\lambda^*_+) > 0	& a_+ = 0    \\
%\mbox{deflagrations:} \dotfill		& \dotfill \\
%  \qquad v_q < c_s             	  	& v_h > v_q  \\
%  \qquad \mbox{Im}(\omega) > 0         	& \mbox{Im}(\omega) > 0    \\
%  \qquad a_1 = 0 	   		& \lambda^*_1 = -\omega/v_0    \\
%  \qquad a_- = 0 		   	& \mbox{Im}(\lambda^*_-) < 0    \\
%  \qquad \mbox{Im}(\lambda^*_+) > 0 	& a_+ < 0    \\
%\hline
%\hline 
%\end{array}
%\nonumber
%\end{equation}

\section{Numerical Results}
\setcounter{equation}{0}
\label{sec:results}

\subsection{Initial Data}
\label{sec:initialdata}

The initial data has just
one free dimension, energy, which we use to specify the
critical transition temperature $T_c$ and normalize it to unity.
It is also convenient to define a characteristic
length scale as the radius of a spherical nucleated
hadron bubble in approximate equilibrium
with an exterior quark plasma. This is estimated by balancing
pressure forces acting from the quark and hadron sides and including
the effect of surface tension, resulting in
$R_{eq} = 2\sigma/(p_h - p_q)$
for the equilibrium radius,
where $\sigma$ is the surface tension, and
$p_h$ and $p_q$ are the pressure fields from the hadron and 
quark sides respectively. For the preliminary one-dimensional
calculations in \S\ref{sec:1dresults}, the computational
box sizes are set to a multiple of the equilibrium
bubble radius $L=xR_{eq}$, with
$x$ generally in the range 20 - 2000.
Typical length scales or box sizes
in these calculations vary from a few hundred  to a few thousand fermi.
By comparison, the proper horizon size at the time of
the QCD phase transition is
\begin{equation}
d_H = 2t_{age} = R(t)\int_0^t dt' / R(t') 
    = \frac{16}{\sqrt{g_*/51.25}~(T_{MeV}/150~\mbox{MeV})^2}~\mbox{km} ~,
\label{eqn:dh}
\end{equation}
or 16 km for 
$g_*=51.25$ and $T_{MeV}=150~\mbox{MeV}$. In writing (\ref{eqn:dh})
we have assumed 
$R(t)$ is the cosmological scale factor in the isotropic FLRW
background model dominated by radiation energy density of the
form $e=a_r T^4 = e_0 R^{-2}$ such that
\begin{equation}
t_{age} = \sqrt{\frac{3 c^2}{32\pi G a_r}}~T^{-2}
        = \frac{2.7\times10^{-5}}
                 {\sqrt{g_*/51.25}~(T_{MeV}/150\mbox{MeV})^2}~\mbox{s}
\label{eqn:tage}
\end{equation}
is the age of the Universe as a function of temperature.

We consider two background temperatures
for the super-cooled initial state triggering the shock
and phase front propagation:
$T=0.9 T_c$ and $T=0.9943 T_c$. 
The latter temperature is chosen to match
the 1+1 dimensional cases considered in reference \cite{IKKL94},
and provides a useful benchmark
against which some of our results can be compared.

To keep the number of numerical simulations down to
a reasonable level, we fix the following additional parameters
for all calculations and for both temperature cases:
critical transition temperature $T_c = 150$ MeV corresponding
to the QCD symmetry breaking energy occurring 
when the Universe was approximately
$t_{age} = 2.7\times10^{-5}$ s old (see equation
(\ref{eqn:tage}));
particle degrees of freedom $g_*= 51.25$ to be consistent
with previous studies \cite{KK86};
surface tension $\sigma = 0.1~T_c$; 
correlation length $\ell_c = 1.0/T_c$; and
latent heat $L=2~T_c^4$.
Note that for both temperatures
considered here, the interaction potential (equation (\ref{eqn:pot})
and Figure \ref{fig:pot}) has two local minima 
(at $\phi = 0$ and $\phi = \phi_{min}(T) > 0$, which allows
for the quark (cold phase with $\phi=0$) and hadron 
(hot phase with $\phi=\phi_{min}(T)$) states to co-exist.

The remaining parameter, the dissipation constant
$\kappa$, is varied over the range
$0.1\le \kappa \le 10$ in the one-dimensional calculations
to determine the regimes in which detonations
and deflagrations are triggered, and to compute the
stability parameters from linear theory.

The scalar field $\phi$ is initialized according to
\cite{IKKL94}
\begin{equation}
\phi(x) = \phi_c (1+\Delta) \frac{2\alpha T \overline{\lambda}}{3\lambda}
          \left(\frac{1-\sqrt{1-f(L-x)}}{f(L-x)}\right) ~,
\end{equation}
with
\begin{equation}
f(x) = \frac{1-\overline{\lambda} \coth^2(\sqrt{\beta (T^2-T_0^2)}~x/(1+\Delta))}
            {1-\coth^2(\sqrt{\beta (T^2-T_0^2)}~x/(1+\Delta))} ~,
\end{equation}
$\overline{\lambda} = 9\lambda \beta (T^2-T_0^2)/(2\alpha^2 T^2)$, and
$x=L$ is the right-most edge along the $x$-axis representing
the computational box size.
These expressions define a hadron phase with $\phi > 0$ at $L-x \rightarrow 0$,
and quark phase with $\phi \rightarrow 0$ as $L-x \rightarrow \infty$,
with an exponentially steep interface boundary of width
set by the model parameters.
The numerical constants $\phi_c$ and
$\Delta = 0.05$ are used to perturb the solution
out of dynamical equilibrium by varying both the amplitude and
width of the field in the hadron phase.
For most of the calculations we set $\phi_c=1$. However, for the
larger $\kappa$ cases we found it useful to
specify $\phi_c$ such that $\phi(x=L) \approx \phi_{min}(T)$, in order
to reduce the transient interval between the initial configuration 
and the time that the phase front fully develops.

\subsection{One-Dimensional Results}
\label{sec:1dresults}

The primary motivations for the one-dimensional calculations
presented here were to narrow the parameter space and help
illuminate the two-dimensional studies presented in the following
section for the different background temperatures considered.
In cases where the temperature was initialized to $T=0.9 T_c$, we used anywhere
from $2 \times 10^3$ to $3.2\times10^4$ zones to 
resolve a length scale of $L=2000 R_{eq} = 2190$ fm.  
For runs in which $T=0.9943 T_c$ we 
used from $5\times10^2$ to $4\times10^3$ zones to
resolve $L= 20 R_{eq} = 452$ fm.  In these 1D parameter studies, 
we also considered an initial background temperature of $T=0.95 T_c$.  
These runs were useful for characterizing some of the intermediate states 
in the 2D runs below.
For these $T=0.95 T_c$ runs we used from $1\times10^3$ to $4\times10^3$ 
zones to
resolve $L= 250 R_{eq} = 1157$ fm.
The higher resolutions in each temperature case
were generally required for the
larger values of $\kappa$ in order to maintain reasonable zone coverage
in the hadron phase, since the bubble growth velocity was sometimes
substantially less than the shock front velocity, or near the transition 
from a detonation front to a deflagration front, as detailed below.  
For this first
group of 1D runs, the code was allowed to
run for approximately one sound-crossing time $t_s=L/c_s$,
where $c_s=1/\sqrt{3}$.

\begin{figure}
\includegraphics[width=10.5cm]{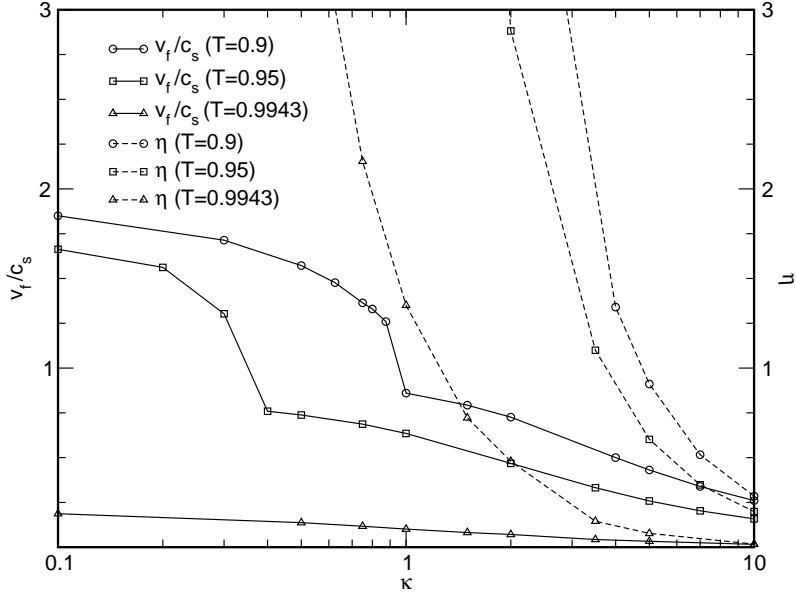}
\caption{Bubble growth velocity normalized to the sound speed
$v_f/c_s$, and stability parameter $\eta$ as a function
of the dissipation constant $\kappa$ for temperatures 
$T/T_c=0.9$, $0.95$, and $0.9943$.  The $T=0.9 T_c$ ($T=0.95 T_c$)
curve switches from a detonation to a deflagration at $\kappa \approx 1$
($\kappa \approx 0.3$).
\label{fig:stability}}
\end{figure}

Figure \ref{fig:stability} shows the phase front or bubble growth velocity
normalized to the sound speed $v_f/c_s$, 
and the quantity $\eta=-T_c v_f dv_f/dT_q$
as a function of the dissipation parameter $\kappa$ for 
temperatures $T/T_c=0.9$, $0.95$, and $0.9943$.
According to reference \cite{HKLLM93}, $\eta$ determines the
stability of bubbles and we compute it using the
approximation $\eta = v_f^2/v_c^2$, where
$v_c^2 = (1/2)(1- T_q^2/T_c^2)$.
Notice the $T=0.9 T_c$ curve switches from a detonation to a 
deflagration at $\kappa \approx 1$ where
$v_f/c_s$ crosses unity, and the $T=0.95 T_c$ curve switches at 
$\kappa \approx 0.3$.  However, the precise transition points are
highly sensitive to the resolution and duration of 
the simulations, and what might appear
as a deflagration at one resolution may
turn out to be a detonation at another.
This is attributed to the solutions approaching
either a Jouget or temporary strong detonation state
(which eventually decays into a weak deflagration) as the
parameter $\kappa$ is increased. In particular,
the post-shock velocity for the $T=0.9 T_c$ case
computed over short time intervals
in the frame of the front decreases from being supersonic
at low values of $\kappa$, to about sonic speed at $\kappa=0.875$,
then subsonic at higher values but less than unity (e.g., $\kappa=0.9$),
corresponding to weak, nearly Jouget, and strong detonations, respectively.
Solutions for which $\kappa$ is close to unity thus represent a regime
in parameter space that does not allow stable detonation modes since strong
detonations are forbidden modes of propagation 
(see, for example, \cite{KL95}).
We note that the conservative hyperbolic form of the hydrodynamics 
equations and
the non-oscillatory central difference methods for solving the
equations generally out-perform artificial viscosity methods in resolving
and maintaining stable evolutions near the critical Jouget state.

Table \ref{tab:defl} summarizes our results for the potentially unstable,
$\eta < 1$, deflagration runs.
$T_h$ and $T_q$ in Table \ref{tab:defl} are the 
average temperatures of the hadron and quark phases, respectively;
and $v_h$ and $v_q$ are the hadron and quark velocities
on either side of the phase front
as measured in the rest frame of the moving front.
The critical wave number, $k_c = (\mu -1)\rho h_q v_q^2 / \sigma$, defines the 
limit above which the system
is stabilized by surface tension \cite{Link92}, where $\mu = v_h/v_q$ and 
$\rho h_q$ is the enthalpy density of the quark phase. 
Modes with wavelengths $\lambda < \lambda_c = k_c^{-1}$ 
are thus expected to be stable.
The instability growth timescale, $\tau$, is estimated as \cite{Link92}
\begin{equation}
\frac{1}{\tau} = \frac{\mu}{1+\mu} \left[ -1 + \left( 1 + \mu - \frac{1}{\mu}
 - \frac{1+\mu}{\mu}\frac{\sigma k_0}{\rho h_q v_q^2} \right)^{1/2} \right] 
k_0 v_q ~,
\end{equation}
where
\begin{equation}
k_0 = \frac{2}{9}\left( \frac{\mu}{1+\mu} \right) 
\left[ 2 + 3\mu - \frac{3}{\mu} - 
\left( 4 + 3\mu - \frac{3}{\mu} \right)^{1/2} \right] \frac{\rho h_q v_q^2}
{\sigma}
\end{equation}
is the wave number of the mode with the shortest growth time scale.  

\begin{table}[tbm]
\caption{
Hadron temperature $T_h$, quark temperature $T_q$, front velocity $v_f$, 
hadron velocity relative to front $v_h$, quark velocity relative to 
front $v_q$, critical stability wavenumber $k_c$, fastest growth 
wavenumber $k_0$, 
linear perturbation growth time $\tau$, and stability parameter $\eta$
as a function of dissipation constant $\kappa$ for 
initial temperatures $T/T_c=0.9$, $0.95$, and $0.9943$.  
The results are
shown only for the potentially unstable 1D deflagration runs
in Figure \protect{\ref{fig:stability}} that satisfy $\eta<1$.
\label{tab:defl}
}
\begin{ruledtabular}
\begin{tabular}{lllllllllll}
$T/T_c$ & $\kappa$ & $T_h/T_c$  & $T_q/T_c$  & $\vert v_{f}/c_s \vert$  & $v_h$
                   & $v_q$ & $k_c$ & $k_0$ & $\tau$ & $\eta$  \\
  \hline
0.9  & 5.0  & 0.90639 & 0.92971 & 0.431 & 0.249
            & 0.193 & 1.816 & 0.933 & $8.25 \times 10^1$ & 0.911 \\
     & 6.0  & 0.90378 & 0.92449 & 0.364 & 0.210
            & 0.164 & 1.250 & 0.642 & $1.46 \times 10^2$ & 0.608 \\
     & 7.0  & 0.90275 & 0.92264 & 0.339 & 0.196
            & 0.153 & 1.071 & 0.550 & $1.83 \times 10^2$ & 0.516 \\
     & 10.0 & 0.89914 & 0.91717 & 0.260 & 0.150
            & 0.118 & 0.613 & 0.314 & $4.17 \times 10^2$ & 0.285 \\
  \hline
0.95 & 5.0  & 0.95436 & 0.96233 & 0.258 & 0.149
            & 0.127 & 0.546 & 0.278 & $6.81 \times 10^2$ & 0.602 \\
     & 7.0  & 0.95246 & 0.95964 & 0.203 & 0.117
            & 0.100 & 0.331 & 0.170 & $1.43 \times 10^3$ & 0.348 \\
     & 10.0 & 0.95072 & 0.95747 & 0.158 & 0.091
            & 0.077 & 0.198 & 0.101 & $3.09 \times 10^3$ & 0.199 \\
  \hline
0.9943 & 1.5 & 0.99633 & 0.99686 & 0.082 & 0.048
             & 0.043 & 0.043 & 0.022 & $4.33 \times 10^4$ & 0.722 \\
       & 2.0 & 0.99596 & 0.99646 & 0.071 & 0.041
             & 0.037 & 0.031 & 0.016 & $7.06 \times 10^4$ & 0.479 \\
       & 3.5 & 0.99527 & 0.99576 & 0.043 & 0.025
             & 0.022 & 0.012 & 0.006 & $2.66 \times 10^5$ & 0.145 \\
       & 5.0 & 0.99491 & 0.99541 & 0.033 & 0.019
             & 0.016 & 0.007 & 0.004 & $6.82 \times 10^5$ & 0.077 \\
       & 10.0& 0.99439 & 0.99492 & 0.016 & 0.009 
             & 0.007 & 0.002 & 0.001 & $5.95 \times 10^6$ & 0.017 \\
\end{tabular}
\end{ruledtabular}
\end{table}

Table \ref{tab:det} summarizes the results computed for the
detonation cases with temperature $T=0.9 T_c$.
Here $v_f$ represents both the front and shock velocities,
and $\rho h_{h}$ is the enthalpy density
of the hadrons immediately behind the phase front.
The quantity $\sigma/\rho h$ has dimensions of length,
and provides a natural scale that gauges the relative
importance of surface tension.
It is also applied as a dimensional scaling variable in the calculations
of reference \cite{Abney94}, which we use as a guide to 
estimate perturbation growth rates and physical run
times for the two-dimensional calculations 
presented in \S\ref{sec:2dresults}.

\begin{table}[tbm]
\caption{
Front velocity $v_f$ and enthalpy to surface tension
ratio $(\rho h_{h})/\sigma$ on the hadron side
as a function of dissipation constant $\kappa$ for 
a selected set of 1D detonations
with initial temperature $T=0.9 T_c$
\label{tab:det}
}
\begin{ruledtabular}
\begin{tabular}{llll}
$T/T_c$ & $\kappa$ & $\vert v_{f}/c_s \vert$  & $(\rho h_{h})/\sigma$ \\
  \hline
0.9  & 0.1  & 1.85 & 191.7  \\
     & 0.3  & 1.71 & 195.7 \\
     & 0.5  & 1.57 & 201.1  \\
     & 0.75 & 1.36 & 216.2  \\
     & 0.8 & 1.33 & 220.4 \\
\end{tabular}
\end{ruledtabular}
\end{table}

\begin{figure}
\includegraphics[width=6in]{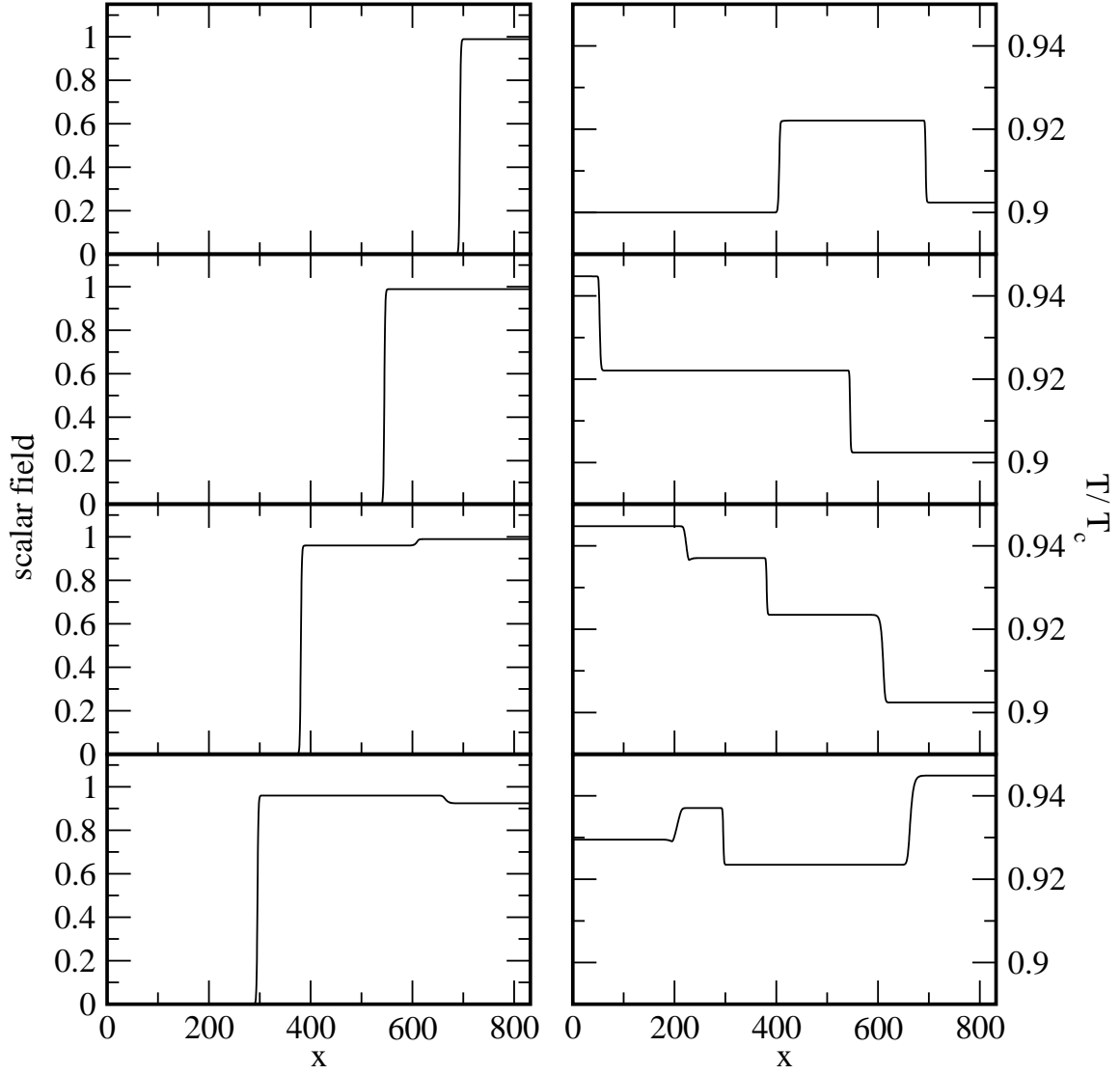}
\caption{Scalar field (left column) 
and temperature (right column)  as a function of position 
for a 1D deflagration
case with $T=0.9T_c$ and $\kappa=7$.  {\it Top row:} Early undisturbed
deflagration and shock fronts at $t=3.2\times10^{-21}$ s.  {\it Second row:}
Shock collision in the quark phase at the left boundary at $t=6.6\times10^{-21}$ s.
{\it Third row:} Reflected shock interacting with deflagration front at 
$t=1.1\times10^{-20}$ s.  {\it Fourth row:} Interaction of two reflected
shocks in the hadron phase at the right boundary at $t=1.4\times10^{-20}$ s.
\label{fig:def-1D_multi}}
\end{figure}

\begin{figure}
\includegraphics[width=6in]{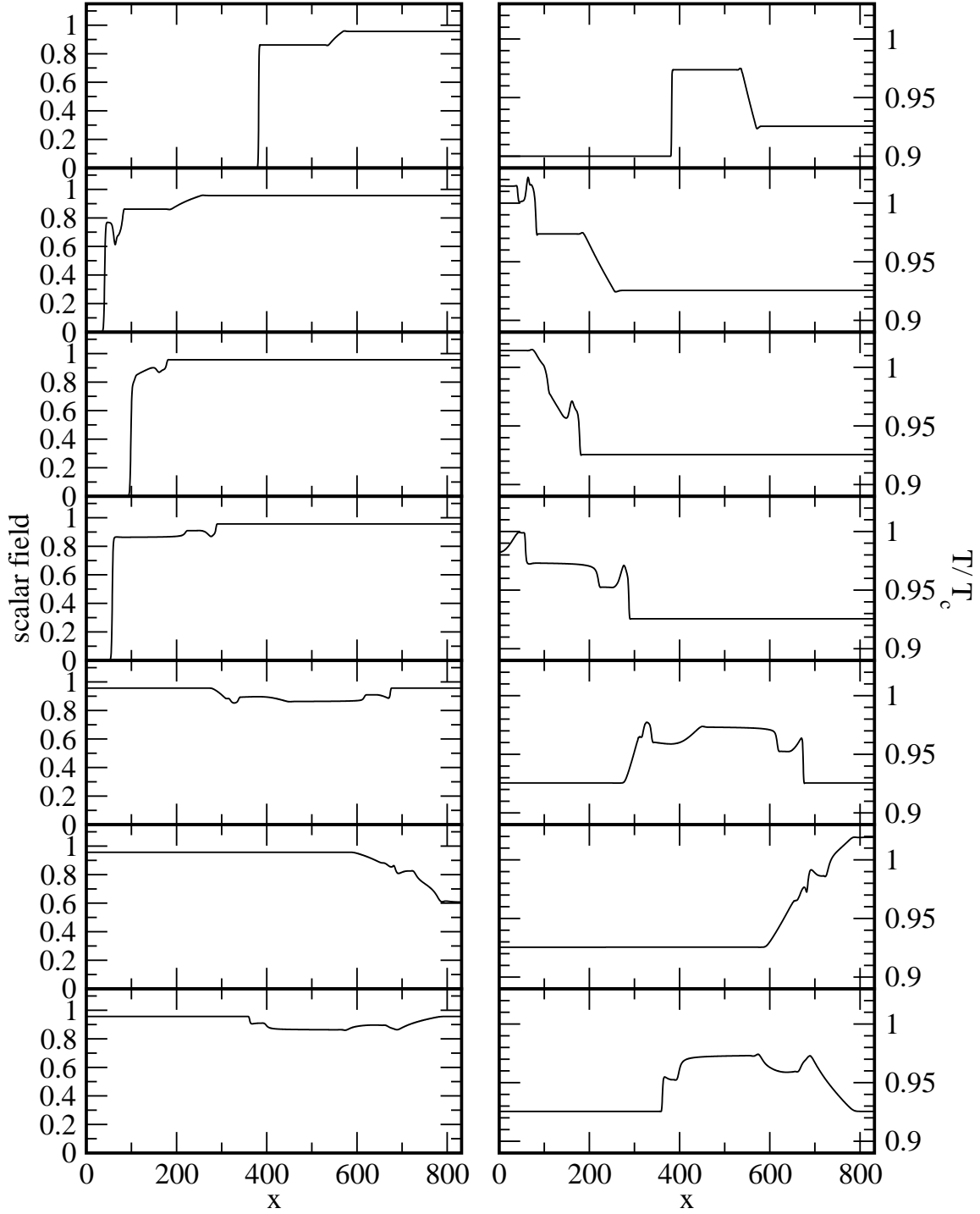}
\caption{Scalar field (left column) and temperature (right column) 
as a function of position 
for a 1D detonation
case with $T=0.9T_c$ and $\kappa=0.5$.  {\it Top row:} Early undisturbed
detonation front plus rarefaction at $t=2.2\times10^{-21}$ s.  
{\it Second row:}
Shock-shock interaction at left boundary at $t=4.8\times10^{-21}$ s.
{\it Third row:} Reflected shock passing through the rarefaction wave at 
$t=5.7\times10^{-21}$ s.  {\it Fourth row:} Interaction of the rarefaction
waves at left boundary at $t=6.6\times10^{-21}$ s.  {\it Fifth row:} 
Reflected state from left boundary at $t=9.7\times10^{-21}$ s.  
{\it Sixth row:}
Interaction of reflected states at right boundary at $t=1.2\times10^{-20}$ s.  
{\it Seventh row:} Reflected state from right boundary at $t=1.5
\times10^{-20}$ s.
\label{fig:det-1D_multi}}
\end{figure}

\begin{figure}
\includegraphics[width=6in]{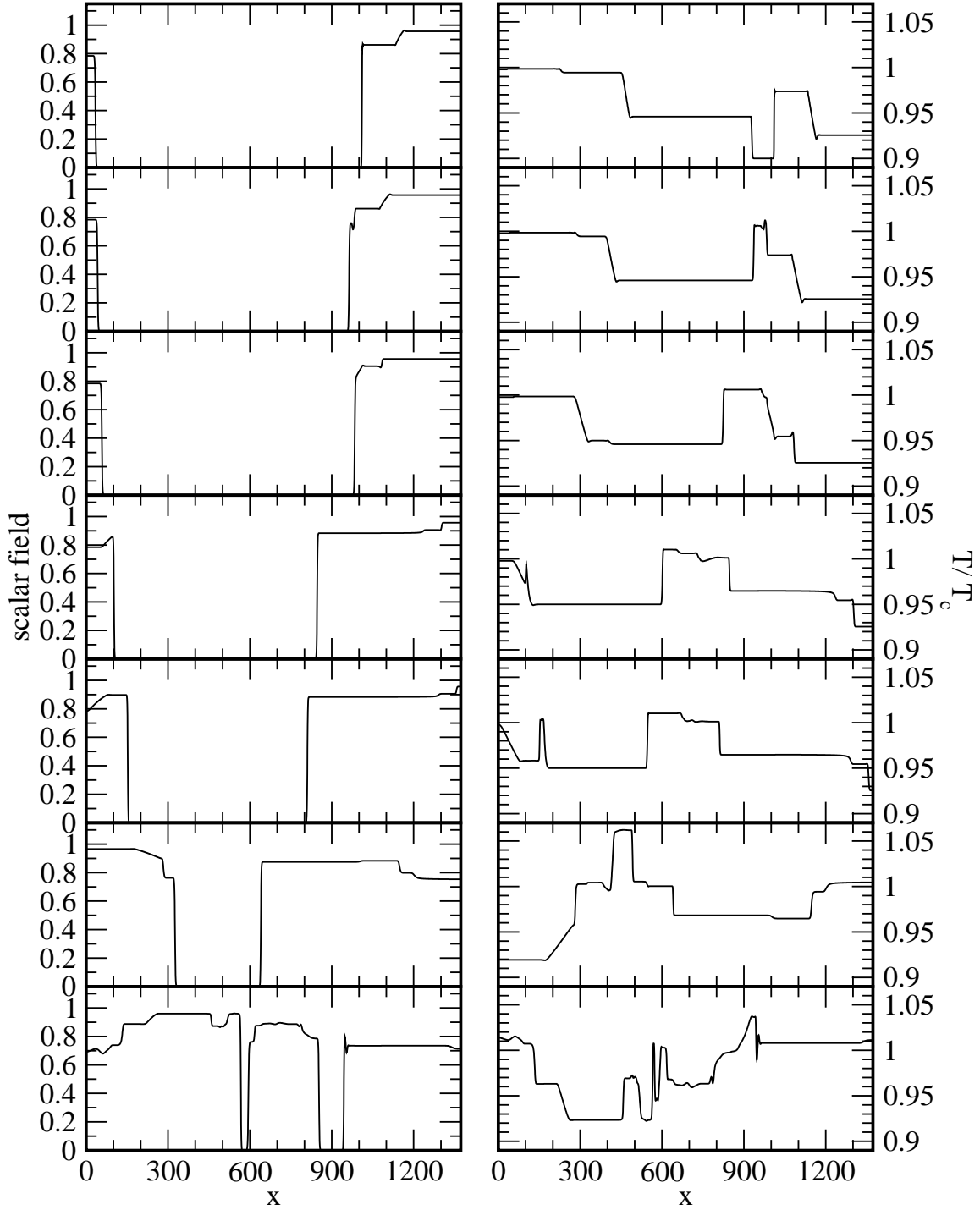}
\caption{Scalar field (left column) and temperature (right column) 
as a function of position 
for a 1D interaction of shocks and phase fronts
originating in two separate regions at different temperatures:
a deflagration system at $T=0.9943 T_c$ at the left end,
and a detonation configuration at $T=0.9 T_c$ at the right end.  
The dissipation constant is
$\kappa = 0.5$ everywhere, and the temperature discontinuity
is initiated at the grid center at $t=0$.  {\it Top row:} Early undisturbed
deflagration and shock fronts (left), detonation front plus rarefaction 
(right), and 
thermal shock front plus rarefaction (center) 
at $t=1.8\times10^{-21}$ s.  {\it Second row:}
Interaction of detonation front and thermal shock at $t=2.2\times10^{-21}$ s.
{\it Third row:} Interaction of the detonation rarefaction and thermal 
shock, and thermal rarefaction and deflagration shock at 
$t=3.1\times10^{-21}$ s.  {\it Fourth row:} Interaction of deflagration 
front and thermal rarefaction, and deflagration shock and
detonation shock at $t=4.8\times10^{-21}$ s.  {\it Fifth row:} 
Acceleration of deflagration front by thermal rarefaction 
at $t=5.3\times10^{-21}$ s.
{\it Sixth row:} Interaction of newly accelerated front and 
detonation front at $t=7.5\times
10^{-21}$ s.  {\it Seventh row:} Interactions of multiple shocks resulting 
in two separate quark regions at $t=1.1\times10^{-20}$ s.
\label{fig:df-dt-1D_multi}}
\end{figure}

In Figures \ref{fig:def-1D_multi}-\ref{fig:df-dt-1D_multi}, we present
more detailed results of three particular 1D runs that will help illuminate
the 2D calculations with similar parameters presented in the next section.
Figure \ref{fig:def-1D_multi} 
shows a series of outputs of the scalar field and temperature for a 
weak (subsonic post-front velocity relative to the phase front) 
deflagration case with $T=0.9T_c$ and $\kappa=7$.  This run
used 2000 zones to resolve a length scale of $L=1095$ fm, and was
allowed to run for slightly longer than two sound-crossing times so that 
the shock leading the deflagration front propagates across the grid twice.
These tiled displays show a leading shock front
(in the temperature graphs) originating
from the right and traveling to the left with velocity 
slightly higher than sound speed $v_s=1.03 c_s$,
heating up the quarks just ahead of the cool hadron
phase ({\it first row}). The deflagration front (separating the two regions 
with different scalar field values) travels to the left at a
speed slower than the leading shock front, $v_{f}/c_s \approx 0.34$.
Due to the reflection boundary conditions imposed on all our
calculations, a mirrored shock collision occurs when
the leading shock hits the left boundary ({\it second row}).
The reflected shock then travels
to the right, eventually colliding with the phase front 
moving in the opposite direction ({\it third row}). 
The shock heats up the fluid
and reduces the scalar field as it passes through
the hadron phase. Upon impact with the phase front, the
shock/front interaction also generates a rarefaction wave 
traveling to the left which cools off the quarks in its path.
Finally the hadron matter at the right end is heated further as the
reflected shock hits the right boundary and collides with another incoming
shock ({\it fourth row}).

Figure \ref{fig:det-1D_multi} shows a similar series of displays
for a weak (supersonic post-shock velocity relative to the phase front) 
detonation case with $T=0.9T_c$ and $\kappa=0.5$,
run for almost two sound-crossing times.  This run
used 2000 zones to resolve a length scale of $L=1095$ fm.  
The outputs begin in the first row 
with an early undisturbed state
showing the leading shock/detonation front originating at the
right end of the grid and traveling to the left
at speed $v_f/c_s \approx 1.6$. A rarefaction wave follows
the shock and separates
two hadron regions at two slightly different temperatures.
The second row shows the shock-shock collision which
occurs when the leading
shock front reaches and reflects off the left boundary.  This interaction 
heats up the fluid to temperatures higher than $T_c$, and
generates a quark region at the left boundary.
As the reflected shock travels to the right, the quark region grows.  
The third row shows the reflected shock passing 
through the rarefaction wave separating the two hadron
states.  This rarefaction wave cools the newly formed 
quark region, converting quarks back into hadrons.
The fourth row shows 
the interaction of reflected rarefaction waves at the left boundary.
The complete reflected state 
showing the thermal distortion and shock profiles after the
entire domain is converted to the hadron phase is displayed in the fifth row.
This compact structure is allowed to traverse the grid and interact
through each sequence again at
the right boundary ({\it sixth row}). 
The complete reflected state from the right 
boundary is shown in the final row.

Figure \ref{fig:df-dt-1D_multi} shows outputs 
from two phase front systems originating in separate 
regions at different temperatures and are thus out of
initial thermal equilibrium: a deflagration front at $T=0.9943T_c$ 
at the left end, and a detonation front at $T=0.9T_c$ at the right end.
This run
used 2000 zones to resolve a length scale of $L=1806.1$ fm.  
The dissipation constant is $\kappa = 0.5$ across both regions, and the 
temperature discontinuity is initiated at the center of the grid.  
The right end is initialized to the same state as the
previous detonation case in Figure \ref{fig:det-1D_multi}.
The left end is similar to the weak deflagration case of
Figure \ref{fig:def-1D_multi}, but
with a smaller dissipation constant and a higher initial temperature.
The first row illustrates an early stage where the deflagration
front has formed at the left end with $v_f/c_s\approx 0.14$,
the shock/detonation front and rarefaction wave have formed at 
the right end with $v_s = v_f \approx 1.6 c_s$, 
and a thermal shock front and rarefaction wave 
are traveling in opposite directions from the 
central thermal discontinuity at about the sound speed.
The second row shows the collision of 
the detonation and thermal shocks.  
The third row shows the passage of the thermal shock
through the detonation rarefaction, and the collision of
the deflagration shock with the thermal rarefaction wave.
The fourth row shows the interaction of the 
deflagration front with the thermal rarefaction at the left
end, and the collision of the deflagration and detonation shocks
at the middle.  
As the rarefaction wave passes through the hadron phase behind
the deflagration front at the left end, the phase transition is accelerated
to near sound speed.
The fifth row shows the accelerated shock and deflagration front features
traveling to the right at velocities $v_s \approx 0.67=1.16c_s$
and $v_f \approx 0.51=0.88c_s$.
Notice that the accelerated front velocity is close to, but slightly 
faster than the deflagration velocity predicted by the 
$T=0.95 T_c$ curve of Figure \ref{fig:stability}, which
yields $v_f = 0.74 c_s = 0.43$.
The sixth row shows the interaction of the newly accelerated front with 
the detonation propagating from the right end.  This interaction 
heats up and decomposes the hadrons back into quarks over
a substantially large region at the center.
The newly formed quark region gradually cools with the
shock passage and rarefaction wave interactions.
Profiles and distributions of quark and hadron regions
become increasingly complex as indicated in the final
row due to the various shock,
phase, and sound wave interactions as they reflect off the
boundaries and propagate through the grid.

We point out that, in general, Figure \ref{fig:stability} cannot always
be used to predict reliably the accelerated velocity or
mode of propagation from rarefaction wave/front interactions.
For example, consider the same initial configuration used in producing
Figure \ref{fig:df-dt-1D_multi}, but with friction parameter $\kappa=0.2$.
In this case Figure \ref{fig:stability} predicts, for a mean field temperature
$T=0.95 T_c$, a supersonic front velocity of $v_f = 1.56 c_s$.
However, the actual (numerical) accelerated solution remains
a weak deflagration with subsonic velocity $v_f = 0.54 = 0.94 c_s$, only
slightly faster than for the $\kappa=0.5$ case.
This holds even in the limit $\kappa \rightarrow 0$:
the accelerated front velocity approaches but does not exceed
the sound speed as $\kappa \rightarrow 0$, and the propagation mode
remains a deflagration with clearly separated phase and shock front features.

\subsection{Two-Dimensional Results}
\label{sec:2dresults}

We focus here on extending to two dimensions several of the
more interesting 
parameter combinations presented in section \S\ref{sec:1dresults}
that may potentially give rise to unstable behavior as
predicted by linear theory. In particular, we consider
six calculations: runs A and B are single deflagrating
fronts; run C is a single detonation transition;
runs D and E are interactions between a planar 
and smaller spherical nucleating bubble of
deflagration (run D) and detonation (run E) types;
and run F is the interaction of a deflagration system with a detonation
front nucleating from two regions out of thermal equilibrium.
Table \ref{tab:2d} summarizes 
the parameters used in each of these runs.

\begin{table}[tbm]
\caption{Summary of parameters for the two-dimensional runs.
\label{tab:2d}
}
\begin{ruledtabular}
\begin{tabular}{llllllll}
Run & $T/T_c$ & $\kappa$ & $\lambda_c$ & $\tau$ 
              & \multicolumn{2}{l}{Grid Dimensions} & Stop Time \\
    &         &          & (fm)        & (s)     
              & (fm) & (zones)                      & (s) \\
  \hline
A & 0.9 & 7.0 & 1.2 & $1.3\times10^{-21}$ & $278.03\times6.14$ 
              & $1280\times64$ & $3.2\times10^{-21}$ \\
B & 0.9943 & 1.5 & 31. & $2.6\times10^{-19}$ & $3059.83\times152.99$ 
              & $1280\times64$ & $7.1\times10^{-20}$ \\
C & 0.9 & 0.5 & 6.6 & $2.2\times10^{-22}$ & $302.34\times32.71$ 
              & $1024\times256$ & $3.5\times10^{-21}$ \\
D & 0.9 & 7.0 & 1.2 & $1.3\times10^{-21}$ & $1000\times1000$ 
              & $512\times512$ & $1.2\times10^{-20}$ \\
E & 0.9 & 0.5 & 6.6 & $2.2\times10^{-22}$ & $1000\times1000$ 
              & $1024\times1024$  & $1.2\times10^{-20}$ \\
F & 0.9943/0.9 & 0.5 & $\dots$ & $\dots$ & $1806.1\times451.53$ 
              & $2048\times512$ & $2.1\times10^{-20}$ \\
\end{tabular}
\end{ruledtabular}
\end{table}

For all of these simulations we initialize the data with various 
perturbations included.  First, the planar fronts in each problem are 
perturbed with a sinusoid of wavelength $\lambda \approx 5 \lambda_c$, 
where $\lambda_c$ is the critical wavelength calculated from our 1D 
studies.  The amplitude of this perturbation is $\delta x/x=0.05$,
and the wavelength $\lambda$ is used to set the 
physical size of the grid parallel to the phase front.  

Transverse and longitudinal 
inhomogeneous fluctuations are also imposed on all the
initial data using the perturbation solutions discussed
in \S\ref{sec:perturbation}.
In particular, $E = E_0 + \delta E$, $v_x = \delta v_x$,
and $v_y = \delta v_y$, with
$E_0 = 3 a_r T^4 + V(\phi,~T) - T \partial_T V(\phi,~T)$,
where
\begin{equation}
\left( \begin{array}{l}
       (\gamma-1) \delta (E/W) \\
       \delta v_x \\
       \delta v_y 
       \end{array}
\right)
=
e^{-i(\omega t + ky)} \sum_j a_j {\bf \widetilde {R}_j}
   e^{\mbox{Im}(\lambda_j^*)x} e^{-i\mbox{Re}(\lambda_j^*)x} ~,
\end{equation}
$\lambda_j^*$ are the eigenvalues,
and ${\bf \widetilde {R}_j} = {\bf {R}_j}/|{\bf {R}_j}|$
are the normalized eigenvectors. The 
expansion coefficients are defined as
$a_j = \mbox{min}(A v_o,~A E_0)$ with 
amplitude constant $A=0.1$ (if they are not set
to zero because of boundedness constraints as discussed
in \S\ref{sec:perturbation}).
$v_0$ is the background average velocity in either
the quark or hadron regions as defined in
\S\ref{sec:perturbation}, but implied here to be 
the velocity component orthogonal
to the interface boundary and measured in the rest frame of
the unperturbed surface. We take
$v_0 = c_s + 0.5(1-c_s)$ or $v_0 = 0.1 c_s$
for the unperturbed velocity in the detonation or deflagration
cases, respectively.
Finally, we also impose random fluctuations in the background temperature
with peak excursions of $\delta T/T = 0.005$.

\subsubsection{Single Front Perturbations}
\label{sec:2d_single}

The growth and/or decay of perturbations are tracked
by calculating the power spectral density (PSD) of various wavelength 
modes as a function of time.  Here the PSD is defined as
$P(k) = 2 \vert C_k \vert^2 / N^2$,
where $N$ is the number of zones parallel to the front, 
$k=1,2,\cdots,(N/2-1)$ is the wavenumber, and $C_k$ is the discrete Fourier 
transform.  We calculate the PSD for the phase front position
as well as the integrated scalar field and integrated 
temperature behind the front as a function of transverse coordinate
(along the $y$-axis).

\begin{figure}
\includegraphics[width=10.5cm]{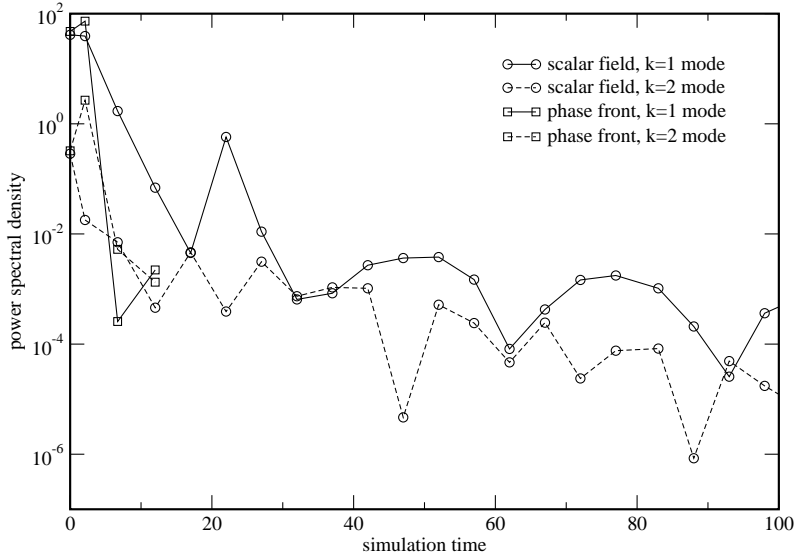}
\caption{Power spectral density as a function of time for the $k=1$ and 
$k=2$ modes of the front position and integrated scalar field behind the 
front for the $T=0.9 T_c$, $\kappa=7$ deflagration case (run A).
\label{fig:psd-def1}}
\end{figure}

Considering the results of section \S\ref{sec:1dresults}, we 
choose for the first deflagration case (run A) the parameters
$T=0.9 T_c$ and $\kappa=7$, which results in a growth timescale 
$\tau$ that is computationally realizable.  We use this
growth time to set the run time 
($t \approx 2.5\tau\approx 3.2\times10^{-21}$ s) 
and the physical length scale ($\sim c_s \tau$) 
perpendicular to the phase 
front.  This problem is resolved with 64 zones parallel 
to the front and 1280 zones perpendicular to the front,
providing similar grid spacing in each of the two dimensions.
Figure \ref{fig:psd-def1} shows the power spectral density as a function of 
time for the $k=1$ and $k=2$ modes of the front position and the integrated 
scalar field behind the front.  The data for the phase front is terminated
at $t\approx 0.03\tau\approx 5\times10^{-23}$ s 
(or $t=12$ in code units),
at which time the perturbations in the front position 
are too small to be resolved at the grid resolution scale.

\begin{figure}
\includegraphics[width=10.5cm]{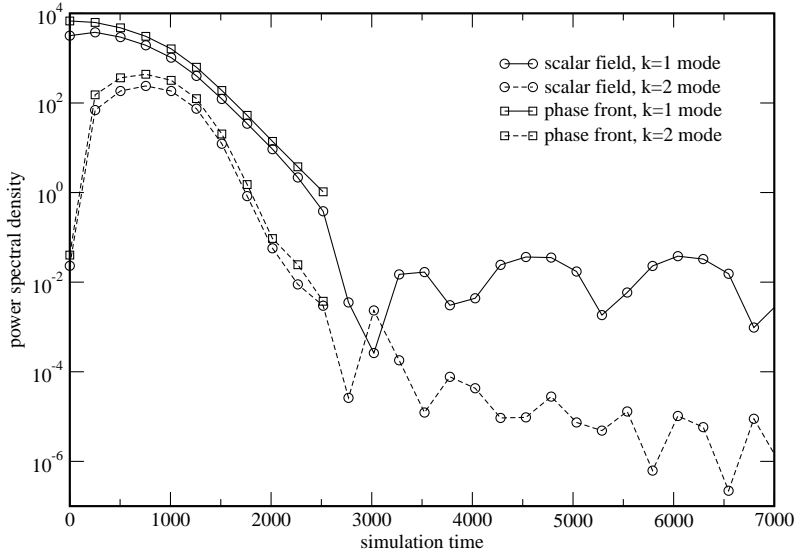}
\caption{Same as Figure \ref{fig:psd-def1}, except for the $T=0.9943 T_c$, 
$\kappa=1.5$ deflagration case (run B).
\label{fig:psd-def2}}
\end{figure}

We also consider a second deflagration 
case with $T=0.9943 T_c$ and $\kappa=1.5$ (run B)
for comparison, despite the fact that the predicted
growth timescale is well beyond what we can
simulate in real time.  Instead we set the physical length scale perpendicular 
to the phase front to be twenty times the scale parallel to the front, which 
is $\approx 5 \lambda_c$ as before.  We then fixed the run time to be 
$t\approx4t_s\approx0.25\tau\approx7.1\times10^{-20}$ s, 
where $t_s=L/c_s\approx100\lambda_c/c_s$ is the sound-crossing time 
perpendicular to the front.  This physical time scale gives 
us a reasonable run time to formulate conclusions about the general 
behavior of perturbations across the shock and phase fronts.
Figure \ref{fig:psd-def2} shows the PSD as a function of 
time for this simulation.  Data for the phase front is terminated at 
$t\approx0.04\tau\approx 10^{-20}$ s ($t=2518$ in code units)
once the perturbations in the front position are smaller than the grid 
resolution.  
Note that the physical scale perpendicular to the front 
for this run is about a factor 
of ten larger than for runs A and C.  This implies that the initial 
perturbation in the phase front is also a factor of ten larger, accounting
for the greater power in each of the modes and partially explaining
the longer decay time.  The decay time is also
influenced by the slower front velocity and longer wavelength in this case.

\begin{figure}
\includegraphics[width=10.5cm]{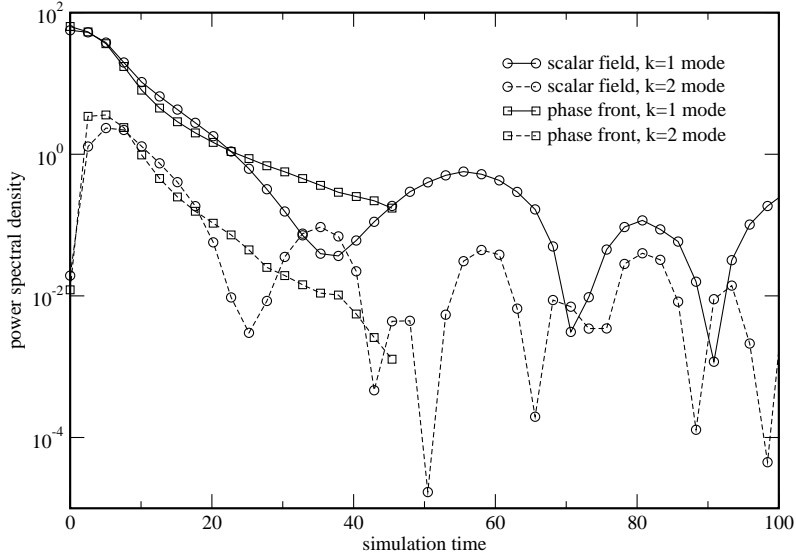}
\caption{Same as Figure \ref{fig:psd-def1}, except for the $T=0.9 T_c$, 
$\kappa=0.5$ detonation case (run C).
\label{fig:psd-det}}
\end{figure}

For the detonation case (run C), we choose
$T=0.9 T_c$, $\kappa=0.5$ and a wavenumber
$k=0.001~(\rho h_h/\sigma)$, which, according to
Figure 2 in \cite{Abney94}, gives a 
growth time of $\tau \sim 1000~(\sigma/\rho h_h)$ for 
Chapman-Jouget detonations. Using the tabulated value
of $\rho h_h/\sigma$ in Table \ref{tab:det}, we set
the simulation run time to $16 \tau = 3.5 \times 10^{-21}$ s.
Figure \ref{fig:psd-det} shows the PSD as a function of time.
Again, data for the planar front position is terminated at
$t\approx \tau$  ($t=45.4$ in code units)
when the perturbations are too small
to be resolved within a single grid cell.

Although the shape function of the phase front and the individual modes
in each simulation
are observed to oscillate in time, the oscillations are strongly damped
over a period that is significantly shorter than the characteristic growth
time predicted by linear theory.
We thus find no evidence of unstable behavior in any of the perturbed
planar deflagration or detonation cases we have investigated, a result
that is generally consistent with the linear analysis and
conclusions of Huet et al \cite{HKLLM93}.

\subsubsection{Multiple Front Interactions}
\label{sec:2d_multiple}

Here we are interested in the question of
whether mixing can occur through a turbulence-type mechanism
triggered by shock or pressure wave distortions across phase boundaries.
For this, we first consider two-dimensional interactions of 
multiple nucleating regions of varying sizes and
radius of curvature.
Runs D and E look at the interaction of a perturbed planar front system with
a smaller nucleating bubble region.  Run D is a deflagration system
with $T=0.9 T_c$, $\kappa=7$, and
run E is a detonation system with $T=0.9 T_c$, $\kappa=0.5$.
The physical dimensions of the grid used in both
problems is $1000 \times 1000$ fm, and is resolved with
$512 \times 512$ cells in run D and $1024 \times 1024$ cells in run E.
In both runs, the small nucleating bubble region has an initial radius of 30 fm,
and the simulations were run to $t\approx 2t_s$, where $t_s$ is
the sound crossing time across the grid.

\begin{figure}
\includegraphics[width=6in]{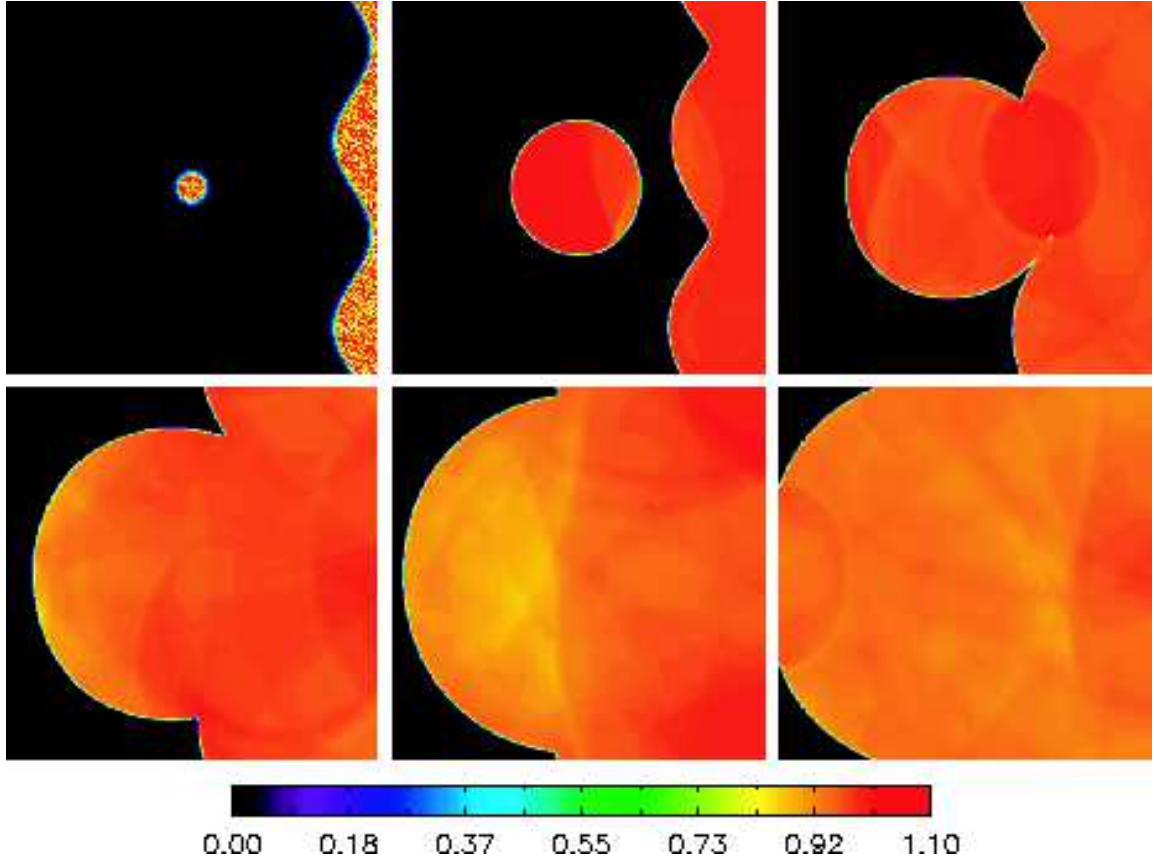}
\caption{
Contour plots of the scalar field for an interacting perturbed plane 
and a smaller nucleating bubble in the deflagration limit
with $T=0.9 T_c$ and $\kappa=7.0$. Results are shown
at times 0, 2.0, 4.1, 6.1, 8.1, and $10.2\times10^{-21}$ s
for grid dimensions of 1000 fm along both the $x$ and $y$ axes.
All images are scaled to the same global color map scheme.
\label{fig:Def+bub_phi}}
\end{figure}

\begin{figure}
\includegraphics[width=6in]{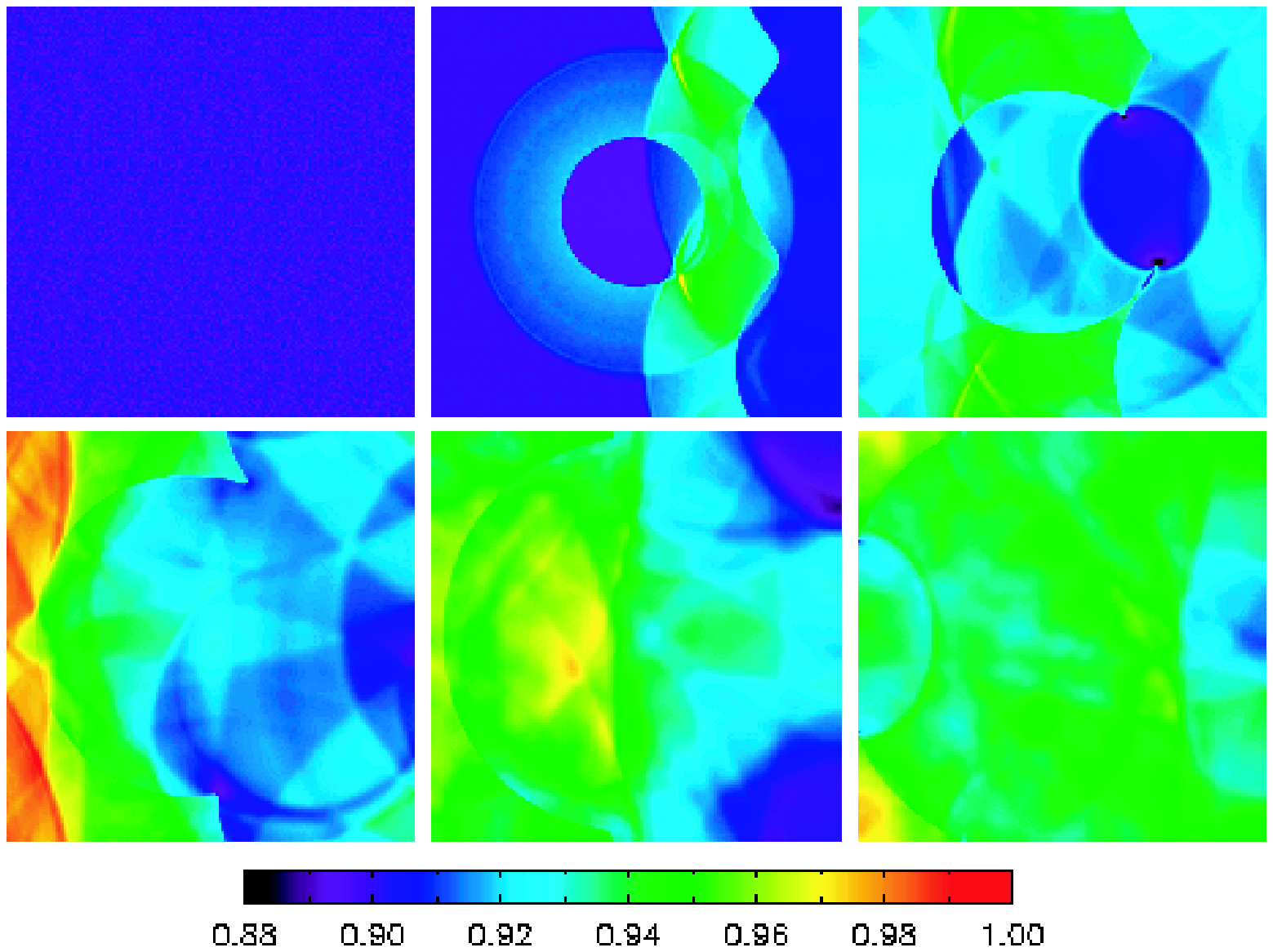}
\caption{
As Figure \protect{\ref{fig:Def+bub_phi}}, but showing the
fluid temperature in units of the critical
temperature ($T_c=1$) at the same corresponding times.
\label{fig:Def+bub_tmp}}
\end{figure}

Figures \ref{fig:Def+bub_phi} and \ref{fig:Def+bub_tmp} show 
the scalar field and the temperature for run D, up to
time $t\approx1.8t_s$ in the final image.
These figures show behavior similar to the one-dimensional plots
of Figure \ref{fig:def-1D_multi}, despite the greater complexity of
the shock and phase geometries, and the additional 
multi-dimensional perturbations. The separation of phase and shock fronts
is clearly evident in the temperature contours, as are the reflected
shock and rarefaction waves generated in the collisions. Perturbations
along and behind the phase fronts are observed to decay in time
as expected from the results of section \ref{sec:2d_single}, and
we observe no signs of turbulent behavior.
In this case, the hadron phases merge smoothly as the bubbles grow, leaving
behind only a spectrum of thermal fluctuations from the initial perturbations,
geometrical scales, oblique shock interactions, and an increasing mean temperature
with each global shock passage.
This scenario is thus consistent with the standard picture of the
QCD transition in which hadron bubbles nucleate and expand as spherical
subsonic condensation discontinuities undergoing a process of
collisions, regular (non-turbulent) coalescence, and quark droplet decay,
to eventually fill the universe.

\begin{figure}
\includegraphics[width=7.5in]{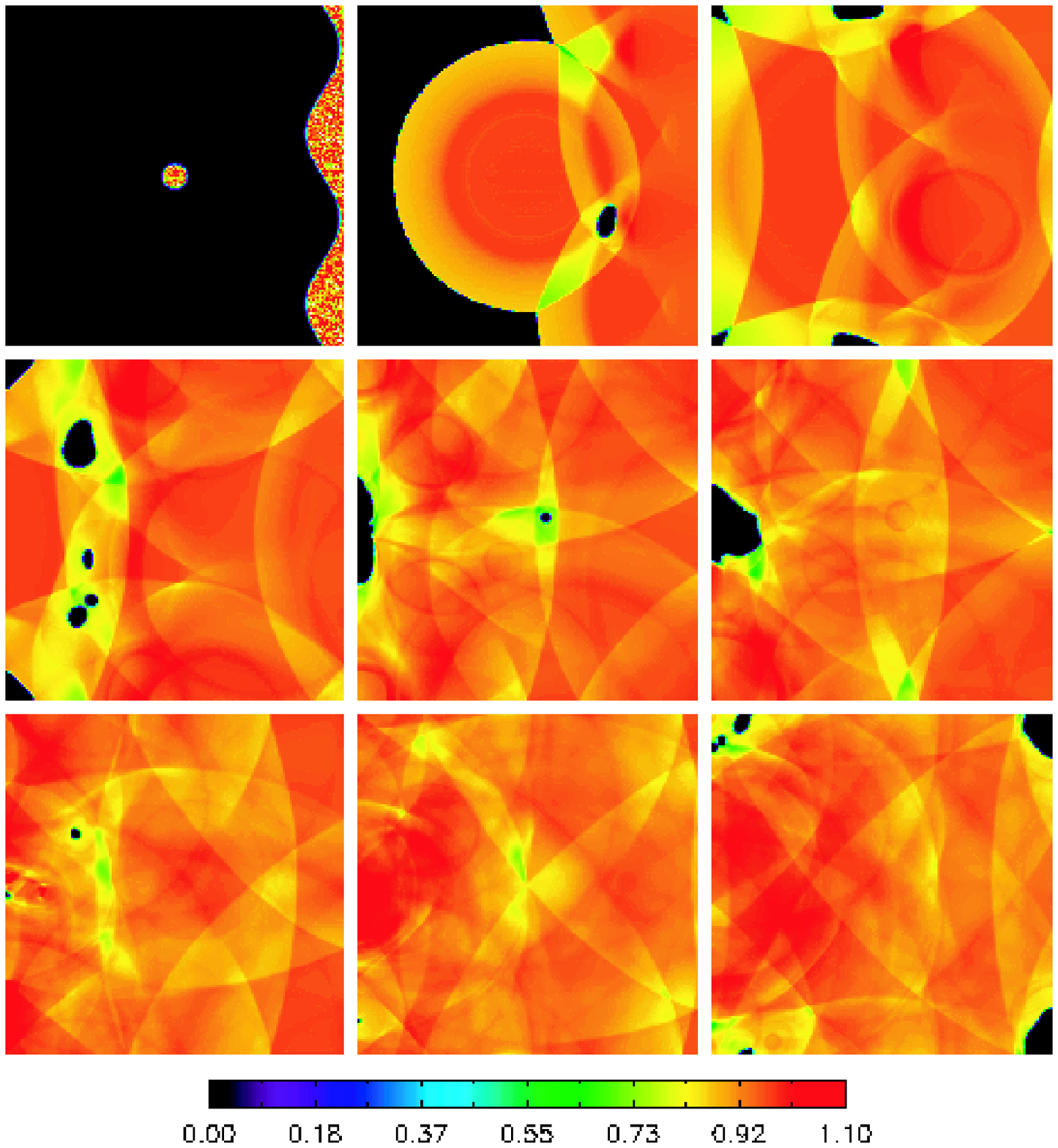}
\caption{
Contour plots of the scalar field for an interacting perturbed plane 
and a smaller nucleating bubble in the detonation limit 
with $T=0.9 T_c$ and $\kappa=0.5$. Results are shown
at times 0, 1.4, 2.7, 4.1, 5.4, 6.1, 7.1, 8.1, and $9.1\times10^{-21}$ s,
on a grid of dimension 1000 fm in each direction.
\label{fig:Det+bub_phi}}
\end{figure}

\begin{figure}
\includegraphics[width=7.5in]{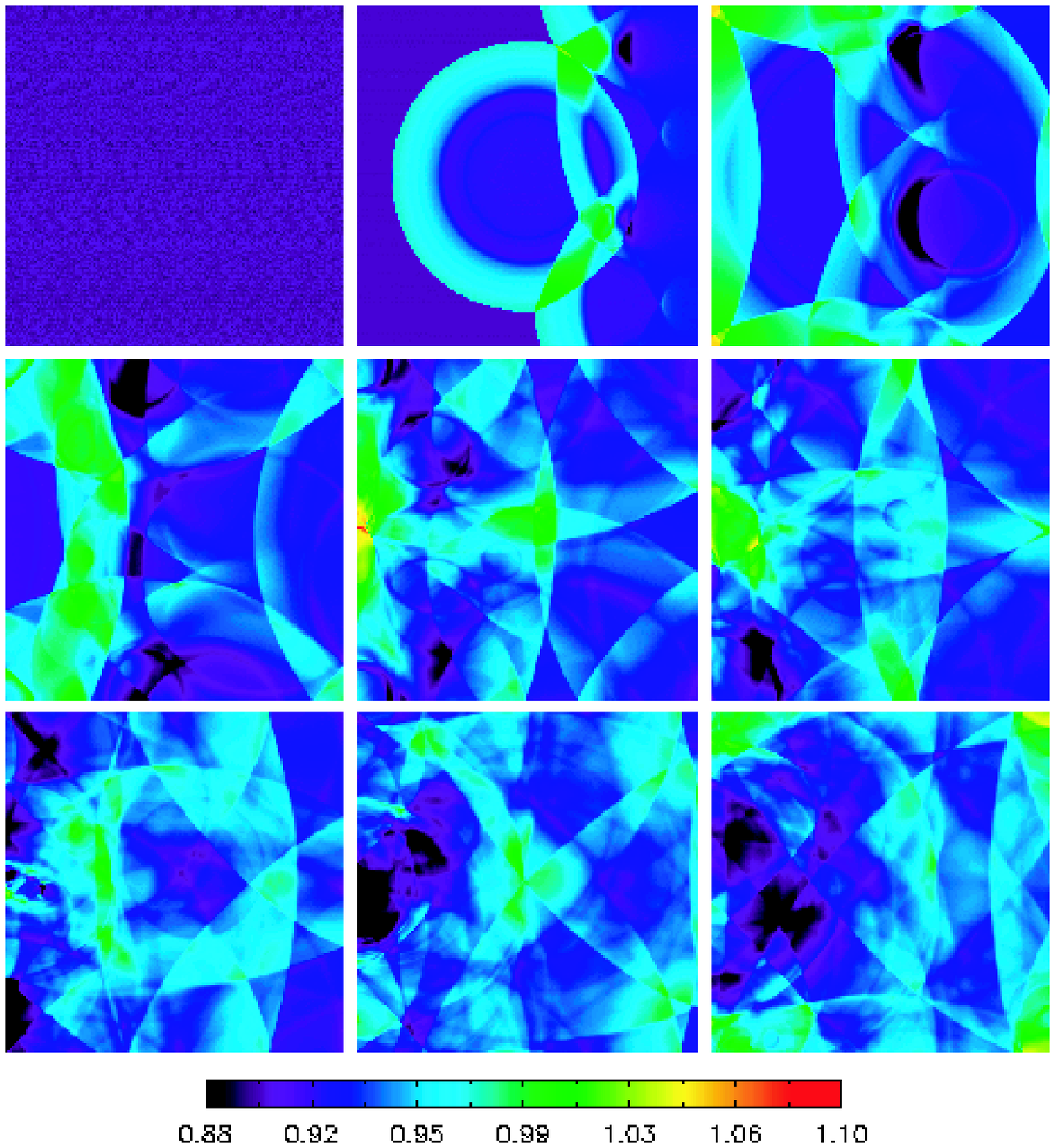}
\caption{
As Figure \protect{\ref{fig:Det+bub_phi}}, but showing the
temperature.
\label{fig:Det+bub_tmp}}
\end{figure}

Figures \ref{fig:Det+bub_phi} 
and \ref{fig:Det+bub_tmp} show the scalar field and temperature for run E
up to time $t\approx 1.6t_s$ in the final image.
In contrast to the deflagration results of Figures \ref{fig:Def+bub_phi}
and \ref{fig:Def+bub_tmp}, these images exhibit more complex behavior.
In particular, front collisions result in the
continuous generation of coherent quark ``nuggets'' formed by the greater
entropy heating of hadrons at shock contact regions, a result consistent
with the one-dimensional calculations shown in 
Figure \ref{fig:det-1D_multi}. The nuggets are generally
formed over spatial scales set by the length of the thermal 
plateau separating the leading shock from the rarefaction wave
(or equivalently by the coherence length of the hot leading phase
of hadrons at the time of impact). 
This, in turn, is set by the mean free path between hadron
bubbles or thermal discontinuities and the strength of
the detonation front which sets the relative velocity
difference between the shock and rarefaction fan.
A simple dimensional argument suggests that maximum quark fragment sizes
are expected to be of order $\delta l = \delta v \delta t \sim L/6$,
or roughly 16\% of the box size. In this crude estimate we assume
for the velocity difference determining the plateau width
$\delta v \sim c_s/3$ as found in the 1D numerical solutions, 
the time between collisions 
$\delta t \sim L/(2 c_s)$, and $L/2$ represents the free
path length between bubbles in this simulation.
The numerical results
find quark regions ranging in size from the smallest resolved
scale of approximately one fermi determined by the cell size, up to a few
hundred fermi, consistent with the above expected result. Once formed,
the nuggets eventually decay away over roughly a sound crossing time
due to adiabatic cooling and the generation of rarefaction waves
at phase boundaries. Nugget shapes are generally determined
by local shock and phase front geometries, but
invariably they tend to become more spherical as the droplets decay
and surface tension effects become significant enough to help erase
surface perturbations.

\begin{figure}
\includegraphics[width=6in]{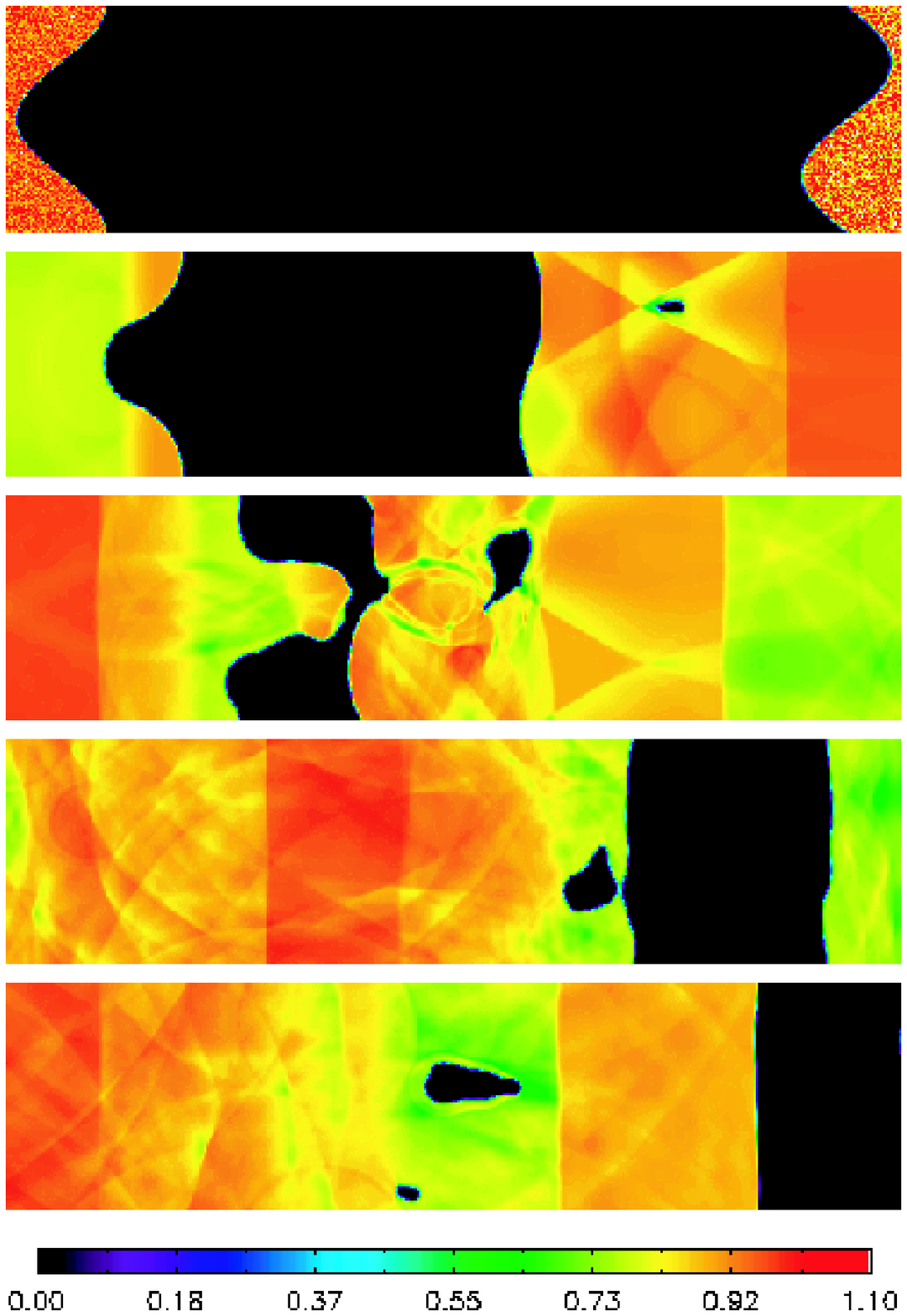}
\caption{
Contour plots of the scalar field for interacting perturbed plane
fronts originating in two separate regions at different temperatures:
a deflagration front at $T=0.9943 T_c$ at the left end,
and a detonation front propagating from the right end
at $T=0.9 T_c$. The dissipation constant is
$\kappa = 0.5$ everywhere on the grid of dimensions
1806.1$\times$451.53 fm, and the temperature discontinuity
is initiated at the grid center along the $x$-axis.
Results are shown at times 0, 4.0, 8.0, 12.5, and $16.5\times10^{-21}$ s.
\label{fig:Det+Def_phi}}
\end{figure}

\begin{figure}
\includegraphics[width=6in]{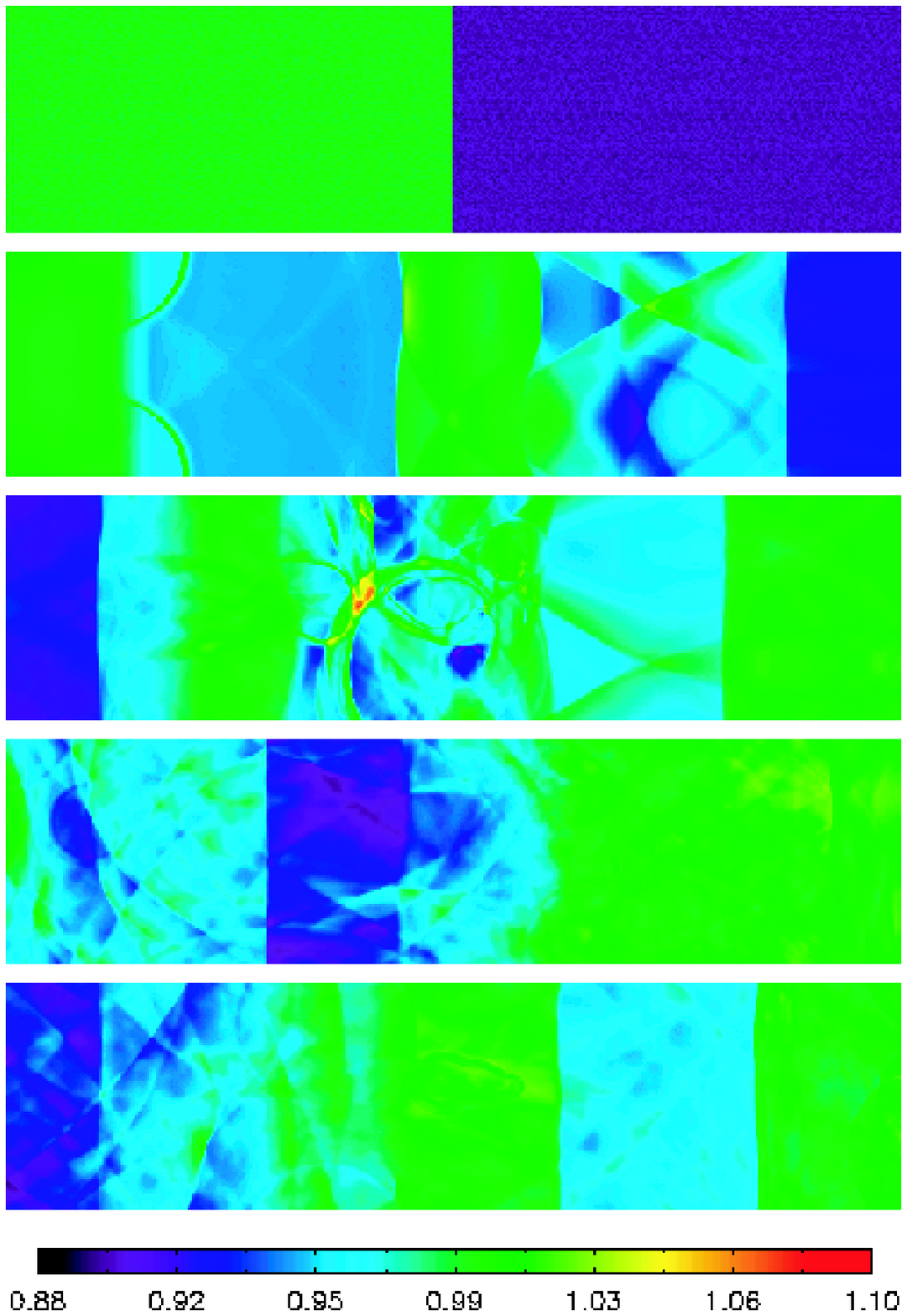}
\caption{
As Figure \protect{\ref{fig:Det+Def_phi}}, but for the fluid temperature.
\label{fig:Det+Def_tmp}}
\end{figure}

A final calculation presented here is run F, which
considers the interaction of a deflagrating system with a detonation
front nucleating from two different regions in space out of thermal equilibrium. 
The parameters are the same as those used in the one-dimensional
calculations of Figure \ref{fig:df-dt-1D_multi} in section \ref{sec:1dresults},
but we also introduce a phase shift of $\pi/2$ between the two fronts
along the transverse direction.
Equal grid spacing is used in the two dimensions here, but with
four times as many zones in the direction perpendicular to the fronts.
The physical size of the grid is set to $1806.1 \times 451.53$ fm
and resolved by $2048 \times 512$ zones, chosen 
as a reasonable balance between resolution and run time.
The physical run time of the final displayed image 
is $t= 1.58 t_s = 1.65\times10^{-20} $ s.
The images in Figures \ref{fig:Det+Def_phi} - \ref{fig:Det+Def_tmp}
for this case show many of the same features exhibited by the
interaction of two detonation fronts, in particular the decomposition
of condensed hadrons into quark droplets or decaying nuggets at shock collisions.
The spontaneously generated nugget regions remain (or become) regular
during their lifetime, as do the phase boundaries of the larger
initial phase fronts. 

Another interesting feature found in this case
is the deflagration `instability' also noted in
the one-dimensional results of Figure \ref{fig:df-dt-1D_multi}.
Although precise quantification of thermal and kinematic
states ahead and behind the accelerated phase front is difficult due 
to the multi-dimensional perturbations, the phase front
is somewhat easier to characterize since perturbations are not as
pronounced in $\phi$.
The deflagration front at the left side, as measured halfway along
the $y$-axis, is accelerated 
from an initial velocity of $v_f = 0.35 c_s = 0.2$ to 
a supersonic speed of about $1.2 c_s = 0.68$
by the passage of a rarefaction wave generated at the thermal
discontinuity. However, the accelerated front velocities measured
along the top or bottom of the $y$-axis is smaller
and closer to the one-dimensional result described in
section \ref{sec:1dresults}. The midpoint of
the phase front, which lags behind the mean front position as shown
in Figure \ref{fig:Det+Def_phi}, thus undergoes a greater acceleration
due to surface tension and mode dissipation effects.
Also, transverse perturbations affect the flow of the hadron
phase by `folding' the cold phase into itself, creating hadron
domains ahead of the main front which enclose decaying
quark regions and eventually merge to
effectively increase the front propagation velocity.

\section{Summary}
\setcounter{equation}{0}
\label{sec:summary}

We have performed both one- and two-dimensional numerical
simulations of first order quark-hadron phase transitions in the
early Universe. The primary goal of these studies was to explore
the nature of the phase transitions beyond linear stability
analysis, and determine if the interface regions between phases
are inheritly stable or unstable when the full nonlinearities
of the relativistic scalar field and
hydrodynamic system of equations are accounted for.
Whether interface boundaries are stable or unstable
can have important consequences in our understanding
of the standard picture of phase transitions
and for the evolving baryon 
number density distribution, which we assume in this paper evolves
hydrodynamically and coupled thermally and kinematically
to a nonlinear scalar field with a quartic self-interaction potential. 
We used results from linear perturbation
theory to define initial fluctuations on either side of 
the single phase front simulations,
and evolved them numerically in time for both weak deflagrations
and detonations. No evidence of unstable
behavior is found in any of the isolated, single perturbed planar 
front cases we considered, despite the fact that the
cases we chose to study are predicted to be unstable
according to several linear analysis calculations. 
Indeed, the power spectra computed
for the phase surface boundary and integrated fields
behind the phase front all suggest that perturbations
decay fairly rapidly, in fact much shorter then
the growth time predicted by perturbation theory.

We are also motivated to
determine whether phase mixing can occur through a turbulence-type
mechanism triggered by shock proximity or the disruption of
interfaces by pressure or shock waves.
To investigate this scenario, we considered three cases:
the first two involve the interaction of perturbed planar fronts
with smaller nucleating hadron bubbles, one being entirely
a deflagration scenario and the other a detonation;
and a third case simulates the interaction between
deflagration and detonation systems arising from
two regions of space which have super-cooled to different
temperatures and are thus out of initial thermal equilibrium.
In the first case of planar-bubble deflagrations, the results
are consistent with the standard picture of cosmological
phase transitions in which
hadron bubbles expand as spherical condensation fronts and undergo a
process of regular (non-turbulent) coalescence, eventually leading
to collapsing spherical quark droplets in a medium of hadrons.
This behavior is generally supported also by the second
case considering the interaction of multiple detonation bubbles.
Although the evolutions in this second case are complicated
by greater entropy heating from shock interactions,
which contributes to the irregular destruction of hadrons and the
creation of quark nuggets, the interfaces
between phases remain coherent. Interface boundaries,
including the original hadron bubbles as well as any newly formed quark nuggets, 
invariably evolve from complex distorted shapes
(influenced by interacting multi-dimensional rarefaction waves, shocks,
and phase fronts) to become more spherical
as the nucleating regions expand, or as the quark
nuggets collapse and surface tension becomes more important.

To summarize, although these calculations exhibit complex behavior,
there are no signs of hydrodynamic mixing instabilities
during this transition period, at least for the choice of parameters,
scalar field interaction potential, equation of state, and
grid resolutions investigated.
Interfaces between quark and hadron phases remain regular,
as perturbations along and across the phase boundaries 
are consistently damped out in time.
Our results are thus generally consistent with the standard
picture of cosmological phase transitions and with the conclusions
of Huet et al. \cite{HKLLM93}, who suggest that
electroweak transitions are stable according to linear
analysis, even if similar definitive statements
can not be made of quark-hadron transitions.

We also note an interesting deflagration `instability' 
or acceleration mechanism evident in
the third case for which we assume an initial thermal
discontinuity in space separating different regions of
nucleating hadron bubbles. The passage of
a rarefaction wave through a slowly propagating deflagration
can significantly accelerate the condensation process
and the phase front to velocities near, or in excess of the sound speed.
This suggests that if the universe super-cools at substantially different
rates within causally connected domains, the dominant
modes of condensation may be through supersonic detonations
or fast moving (nearly sonic) deflagrations,
assuming that conditions
for triggering the instability are typical. 
A similar speculation was made by
Kamionkowski and Freese \cite{KF92} who suggested that
deflagrations become unstable to perturbations
and are converted to detonations
over scales determined by linear theory and surface tension.
In this scenario, instabilities distort the bubble shape,
thereby increasing the surface area of the wall which
accelerates the transition by increasing the rate of condensation
and the effective velocity of expansion. However,
for the calculations presented here, deflagrations are accelerated
predominately not from turbulent mixing and surface distortion, 
but from enhanced super-cooling by rarefaction
waves generated across thermal discontinuities.
This effect can be significantly more pronounced in the presence
of both longitudinal and transverse perturbations as we found
by comparing one and two dimensional calculations with the
same initial mean thermal states.
In higher dimensions, the acceleration mechanism is
exaggerated further by upwind phase mergers due to transverse
flow, surface distortion, and mode dissipation effects, a
combination that can result in supersonic front propagation speeds.

\begin{acknowledgments}
This work was performed
under the auspices of the U.S. Department of Energy by
University of California, Lawrence
Livermore National Laboratory under Contract W-7405-Eng-48.
\end{acknowledgments}

\bibliography{references}

\end{document}